\documentclass[12pt]{article}
\pdfoutput=1
\usepackage{amsmath,amssymb}
\makeatletter

    \@addtoreset{equation}{section}
\makeatother
\usepackage[pdftex]{graphicx}

\newcommand{\One}{{\hbox{{\rm 1{\hbox to 1.5pt{\hss\rm1}}}}}}

\newcommand{\CL}{{\cal L}}
\newcommand{\CO}{{\cal O}}

\newcommand\be{\begin{equation}}
\newcommand\ee{\end{equation}}
\newcommand\bea{\begin{eqnarray}}
\newcommand\eea{\end{eqnarray}}
\newcommand\ba{\(\begin{array}}
\newcommand\ea{\end{array})\ }

\begin{document}
\thispagestyle{empty}
\vspace*{2cm}
\begin{center}
 {\LARGE {Matrix models for irregular conformal blocks and\\[4mm] Argyres-Douglas theories}}
\vskip2cm
{\large 
{Takahiro Nishinaka$^{a,}$\footnote{email: nishinaka\_t~[at]~me.com}
}
and\hspace{2mm} Chaiho Rim$^{b,}$\footnote{email: rimpine~[at]~sogang.ac.kr}
}
\vskip.5cm
{\it Department of Physics$^{\,b}$ and Center for Quantum Spacetime (CQUeST)$^{\,a,b}$
\\
Sogang University, Shisu-dong, Mapo-gu, Seoul 121-742 Korea}
\end{center}
%
%
%
 \vskip2cm
\begin{abstract}
As regular conformal blocks describe the $\mathcal{N}$=2 superconformal gauge theories in four dimensions,  irregular conformal conformal blocks are expected to reproduce the instanton partition functions of the Argyres-Douglas theories.
In this paper, we construct matrix models which reproduce the irregular conformal conformal blocks of the Liouville theory on sphere, by taking a colliding limits of the Penner-type matrix models. The resulting matrix models have not only logarithmic terms but also rational terms in the potential. We also discuss their relation to the Argyres-Douglas type theories.
\end{abstract}
%
%
\newpage
\tableofcontents

\section{Introduction} 

The Liouville conformal blocks on Riemann surfaces are conjectured to be equal to the Nekrasov's instanton partition functions of four-dimensional $\mathcal{N}=2, SU(2)$ superconformal quiver gauge theories \cite{AGT:2009}, which is now referred to as the AGT relation. The complex structure moduli of the (punctured) Riemann surface are identified with the marginal gauge couplings of the four-dimensional theory, and the external momenta of the vertex operators are now encoded in the mass parameters of hypermultiplets. The conformal dimensions of the intermediate states are then interpreted as the Coulomb branch parameters of the gauge theory.

After the above conjecture, the notion of ``irregular conformal blocks'' was given in \cite{G:2009} in order to generalize the AGT relation to asymptotically free gauge theories. The asymptotically free theories are obtained by taking a scaling limit of mass parameters while tuning the gauge couplings so that the physical quantities of the low energy theory remain finite. Such a limit corresponds to a colliding limit of two regular vertex operators in the Liouville conformal block, which gives rise to an irregular state, or an irregular vertex operator in CFT. The irregular states constructed in \cite{G:2009} are eigenstates of $L_1$ or $(L_1,L_2)$ with non-vanishing eigenvalues. Their generalization to eigenstates of $(L_n,\cdots,L_{2n})$ with non-vanishing eigenvalues was recently given in \cite{GT:2012} by considering a colliding limit of many regular vertex operators (See also \cite{Bonelli:2011aa}). Since such a general colliding limit gives Argyres-Douglas theories in the gauge theory side, it was pointed out that the conformal blocks with general irregular states inserted should reproduce the instanton partition functions of the Argyres-Douglas theories \cite{Bonelli:2011aa, GT:2012}.

On the other hand, the Liouville {\it regular} conformal blocks are known to be written as the $\beta$-ensemble of matrix models with logarithmic potentials \cite{Dijkgraaf:2009pc}, where the Liouville charge is encoded in the $\beta$-deformation parameter of the matrix model. The external momenta of the Liouville vertex operators are now interpreted as parameters in the matrix model potential and the matrix size $N$, while the conformal dimensions of the intermediate states are encoded in filling fractions, or eigenvalue distributions, of the matrix model. This observation is based on some earlier works on the conformal symmetry hidden in matrix models \cite{Marshakov:1991gc, Kharchev:1992iv, Morozov:1995pb, Kostov:1999xi}. From the AGT viewpoint, this type of matrix models are expected to reproduce the Nekrasov partition functions of $\mathcal{N}=2,\, SU(2)$ superconformal linear quiver gauge theories, and there have been various studies in this direction \cite{Itoyama:2009sc, Eguchi:2009gf, Schiappa:2009cc, Mironov:2009ib, Fujita:2009gf, Itoyama:2010ki, Mironov:2010ym, Morozov:2010cq, Eguchi:2010rf, Itoyama:2010na, Maruyoshi:2010pw, Bonelli:2010gk, Itoyama:2011mr, Nishinaka:2011aa, Galakhov:2012gw, Bourgine:2012gy}.

In this paper, we study the matrix model side of the colliding limit of the Liouville vertex operators, and derive a series of matrix models which reproduce the general irregular conformal blocks of the Liouville theory on sphere. The colliding limit leads to not only logarithmic terms but also {\it rational} terms in the matrix model potential. In fact, we argue that {\it if the matrix model potential is written as a sum of logarithmic and/or rational terms then its partition function reproduce a Liouville conformal block with insertions of regular and/or irregular vertex operators.} Some of the simplest cases were already studied in \cite{Eguchi:2009gf} and expected to reproduce the instanton partition functions of $SU(2)$ gauge theories with two and three flavors. On the other hand, in this paper, we conjecture that other general matrix models with logarithmic and/or rational potentials reproduce the instanton partition functions of Argyres-Douglas theories as well as some asymptotically free theories involving the Argyres-Douglas theories as their building blocks.

The rest of this paper is organized as follows. In section \ref{sec:Liouville}, we review the colliding limit of regular vertex operators and the property of the irregular states. In section \ref{sec:Penner_and_irregular}, we illustrate how the same colliding limit in the matrix model side gives rise to a potential with logarithmic and rational terms. We also specify matrix model potentials for general irregular conformal blocks. Among the series of matrix models we have constructed in section \ref{sec:Penner_and_irregular}, we study the $D_{2n}$ type matrix models in detail in section \ref{sec:irregular_partition}. We demonstrate that the eigenvalues of $(L_n,\cdots,L_{2n})$ for the irregular states are interpreted as parameters of the matrix model potential while the actions of $L_k$ for $0< k< n$ are now encoded in the filling fractions of the matrix model. In section \ref{sec:irregular_partition}, we also illustrate how our matrix models reproduce the small parameter behavior of an irregular state of rank 2. In section \ref{sec:gauge_theories}, we discuss the relation between our matrix models and Argyres-Douglas type theories, by using the Hitchin system with irregular singularities. We show that our matrix models correctly reproduce the Seiberg-Witten curve of the corresponding gauge theories.

\section{CFT and irregular singularity}
\label{sec:Liouville}

In \cite{G:2009}, the notion of an ``irregular vector'' in the representation of the Virasoro algebra was given in the context of the AGT relation \cite{AGT:2009}. It was defined as the following simultaneous eigenstate $|I^{(1)}\rangle$ in the Verma module for a highest weight state of conformal dimension $\Delta_\alpha= \alpha(Q-\alpha)$:
\begin{eqnarray}
L_1|I^{(1)}\rangle = \Lambda_1|I^{(1)}\rangle,\qquad L_2 |I^{(1)}\rangle = \Lambda_2|I^{(1)}\rangle,\qquad L_{n\geq 3}|I^{(1)}\rangle = 0.
\end{eqnarray}
The explicit expression for $|I^{(1)}\rangle$ was obtained in \cite{Marshakov:2009gn}. The inner product of this state is known to be equal to the instanton partition functions of some asymptotically free $\mathcal{N}=2,\,SU(2)$ gauge theories \cite{G:2009, Hadasz:2010xp}.
The generalization of this irregular vector to more simultaneous eigenvalues was given in \cite{GT:2012} as follows (See also \cite{Bonelli:2011aa}).
First, note that the subalgebra
\be
\left[ L_k, L_m \right] = (k-m) L_{k+m} ~~{\rm for ~} k, m >0
\label{vir_pos}
\ee
of the Virasoro algebra implies that a simultaneous eigenstate of $L_k$ and $L_m$ is also an eigenstate of $L_{k+m}$ with a vanishing eigenvalue. Therefore, when we consider a vector satisfying
\begin{eqnarray}
L_k |I^{(n)} \rangle = \Lambda_k |I^{(n)}\rangle \qquad {\rm for}\qquad n\leq k\leq 2n
\label{eq:condition}
\end{eqnarray}
with non-vanishing eigenvalues $\Lambda_k$, we find $L_{k}|I^{(n)}\rangle = 0$ for $k>2n$. Note that the action of $L_{k}$ for $k<n$ is not diagonalized by $|I^{(n)}\rangle$.
This vector $|I^{(n)}\rangle$ is called an irregular vector of rank $n$, which depends on the collection of non-vanishing eigenvalues  $\Lambda = \{ \Lambda_n, \cdots, \Lambda_{2n}\} $. 

It is noted that the irregular vector may arise when one considers so-called collision limit 
of many  primary fields, a certain limiting process which put many fields at the same point. 
Let us consider a state obtained from $(n+1)$ primary fields,  
$|R_n \rangle = \prod_{k=0}^{k=n} \Psi(y_k; \Delta_k) |0\rangle$.
This state satisfies the operator product expansion
\be
T(z) |R_n \rangle = \sum_{r=0}^{n}  
\Big( 
\frac {\Delta_r}{(z-y_r)^2} + \frac1 {z-y_r} \frac \partial {\partial y_r} + {\rm regular ~ terms}
\Big) 
 |R_n \rangle
\label{eq:regular_primary}
\ee
where $ \Delta_r = \alpha_r (Q-\alpha_r)$ is the conformal dimension of the primary field $\Psi(y;\Delta_r)$.  
Taking the limit $y_r \to 0,\,\alpha_r\to \infty$
while keeping
\begin{eqnarray}
c_0 = \sum_{r=0}^n \alpha_r,\qquad c_k =  \sum_{r=0}^n \,\sum_{0 \le s_1 < \cdots < s_k \le n\,(s_i\neq r)}\!\!\! \alpha_r\, \prod_ {i=1}^k  (-y_{s_i})
\end{eqnarray}
finite, one may have   
\be
T(z) |R_n \rangle  \to    
\Big(  
\sum_{k=n}^{2n} 
\frac{\Lambda_k}{z^{k+2}} 
+ 
\sum_{k=0}^{n-1} 
\frac{\CL_k}{z^{k+2}} 
+ 
\frac{L_{-1}} {z} 
+ {\rm regular ~ terms}
\Big) 
 |I^{(n)} \rangle\,,
\label{eq:T_irregular}
\ee
where $T(z) = \sum_{k} L_k/z^{k+2}$ and $|I^{(n)}\rangle \equiv \lim_{y_r\to 0,\, \alpha_r\to\infty}\big(\prod_{0\leq r<s\leq n}(y_r-y_s)^{2\alpha_r\alpha_s}|R_n\rangle\big)$.
 Note here that our $c_0$ is denoted by $\alpha$ in \cite{GT:2012}, and we sometimes write $\alpha$ as $c_0$.
The constants 
\begin{eqnarray}
\Lambda_k= (k+1)Qc_k - \sum_{\ell=0}^{k}c_\ell c_{k-\ell}
\end{eqnarray}
 are eigenvalues of $L_k$, where we set $c_\ell \equiv 0$ unless $0\leq \ell \leq n$. On the other hand, $\CL_k$ is of a certain operator satisfying the Virasoro algebraic relation (\ref{vir_pos}), whose explicit expression is given by
\begin{eqnarray}
\CL_k = \Lambda_k + \sum_{\ell\in \mathbb{N}}(\ell-k)
c_{\ell}\frac{\partial}{\partial c_{\ell-k}}.
\label{eq:op-CL}
\end{eqnarray}
Thus, the above colliding limit leads to an irregular vector $|I^{(n)}\rangle$ of rank $n$ satisfying
\begin{eqnarray}
L_{k}|I^{(n)}\rangle = \left\{
\begin{array}{l}
0 \qquad\qquad {\rm for}\quad 2n+1\leq k,\\
\Lambda_k|I^{(n)}\rangle \quad {\rm for} \quad n\leq k\leq 2n,\\
\mathcal{L}_k|I^{(n)}\rangle \quad {\rm for}\quad 0\leq k\leq n-1.\\
\end{array}
\right.
\end{eqnarray}

Note that $(n+1)$ eigenvalues $\Lambda_n,\cdots,\Lambda_{2n}$ are now encoded in $\alpha,c_1,\cdots,c_n$. The coefficients $c_1,\cdots, c_n$ are identified with the eigenvalues of coherent states in the free field construction of the irregular states \cite{GT:2012}. For $n\geq 2$, the quantities $\alpha,c_1,\cdots,c_n$ are not enough to determine the irregular vector $|I^{(n)}\rangle$ uniquely. In order to fix such an ambiguity, it was proposed in \cite{GT:2012} that an irregular vector of rank $n\geq 2$ can be recursively constructed from irregular vectors with lower ranks. For example, $|I^{(2)}\rangle$ is proposed to be expanded as
\begin{eqnarray}
|I^{(2)}(c_1,c_2;\alpha)\rangle = c_2^{\nu_2}c_1^{\nu_1}e^{(\alpha-\alpha')\frac{c_1^2}{c_2}}\sum_{k=0}^\infty c_2^k|I^{(1)}_{2k}(c_1,\alpha')\rangle,
\label{eq:ansatz}
\end{eqnarray}
where
\begin{eqnarray}
\nu_1 = 2(\alpha-\alpha')(Q-\alpha'),\qquad \nu_2 = (\alpha-\alpha')\left(-\frac{3}{2}Q + \frac{3}{2}\alpha' + \frac{1}{2}\alpha\right).
\end{eqnarray}
Here $|I^{(1)}_{2k}(c_1;\alpha)\rangle$ are so-called ``generalized descendants'' which are linear combinations of vectors obtained by acting $L_{-k}$ and $c_1$-derivatives on $|I^{(1)}(c_1;\alpha)\rangle$. In particular, $|I^{(1)}_0(c_1;\alpha)\rangle = |I^{(1)}(c_1;\alpha)\rangle$ which can be uniquely determined by $c_1$ and $\alpha$. It follows from \eqref{eq:ansatz} that the rank 2 irregular vector $|I^{(2)}\rangle$ depends on an additional parameter $\alpha'$ in addition to $\alpha,c_1,c_2$. The origin of $\alpha'$ can be understood when we note that the regular vector $|R_{2}\rangle$ originally depends on the conformal dimension of the intermediate state.

By generalizing the above argument, we can consider conformal blocks with many irregular vertex operators inserted, which are called the ``irregular conformal blocks.''

\section{Penner model and irregular singularity} 
\label{sec:Penner_and_irregular}

In this section, we derive matrix models which reproduce the irregular conformal blocks of the Liouville theory. We start with the matrix model for regular conformal blocks \cite{Dijkgraaf:2009pc} and take the same colliding limit in the matrix model side.

\subsection{Penner type matrix models for regular conformal blocks}

Let us first consider the $(n+2)$-point correlation function of the Liouville field theory
\be
\left \langle e^{2 \alpha_\infty\phi(\infty)} 
\, e^{2 \alpha_0 \phi (0)} \,\prod_{k=1}^n e^{2 \alpha_k \phi(z_k)} \right\rangle
\label{eq:correlator}
\ee
which is given by the product of holomorphic and anti-holomorphic regular conformal blocks. 
One may evaluate the correlation explicitly in perturbative expansion of the Liouville theory using the free correlation 
$
\langle e^{2 \alpha_1 \phi(z)}  e^{2 \alpha_2 \phi(w)}  \rangle 
= |z-w|^{-4 \alpha_1 \alpha_2}\,. 
$
The correlation is not vanishing if the Liouville charges satisfy the  neutrality condition
\be
\sum_{i=1}^n \alpha_i +\alpha_\infty+ \alpha_0 + bN = Q
\label{eq:screeningcondition}
\ee
where $N$ is the number of the screening charges so that
the correlation is given by
\be
\left\langle  \left(\prod_{I=1}^N\int d\lambda_Id\overline{\lambda}_I\; e^{2b\phi(\lambda_I)}\right)
e^{2 \alpha_\infty\phi(\infty)} 
\, e^{2 \alpha_0 \phi (0)} \,\prod_{k=1}^n e^{2 \alpha_k \phi(z_k)} \right\rangle_{\!\! \text{Coulomb gas}}.
\ee 

The holomorphic part of the correlation, or conformal block, is then written as
\begin{eqnarray}
Z_{\beta-{\rm Penner}}\prod_{0\leq k<\ell \leq n} (z_k- z_\ell)^{-2 \alpha_k \alpha_\ell}
\end{eqnarray}
up to divergent prefactor, where $Z_{\beta-{\rm Penner}}$ is the partition function of the following $\beta$-deformed matrix model \cite{Dijkgraaf:2009pc}
\begin{eqnarray}
Z_{\beta-{\rm Penner}} \equiv   \int  \left[ \prod_{I=1}^Nd\lambda_I \right]
 ~ \Delta(\lambda)^{2\beta} ~\exp\left(\frac{\sqrt{\beta}}{g} \sum_I V(\lambda_I)\right).
\label{eq:penner}
\end{eqnarray}
Here $\Delta(\lambda)=\prod_{1\leq I<J\leq N}\left(\lambda_I-\lambda_J\right),\,\sqrt{\beta}=-ib$, and the potential are given by
\be
V(z) =  \hbar\left[\alpha_0 \log z+ \sum_{k=1}^n  \alpha_k \log (z-z_k)\right],
\label{eq:pot0}
\ee
with $\hbar\equiv -2ig$.
We treat $\hbar$ as a scaling parameter which relates the parameters of the Liouville theory with those of the matrix model.
When $\beta=1$, the integral \eqref{eq:penner} reduces to the Penner-type matrix model, where $\lambda_I$ is regarded as the eigenvalue of a hermitian matrix. 
The matrix model integral is not well defined unless the integration range  
and the Liouville parameter $b$ are to be appropriately assigned.  
To fix this problem, one may consider the integrals with analytical continuation 
and put $b^2= -\beta^2$ (or $b = i\sqrt{\beta}$) 
so that integration is well defined even when any two of the integration variables coincide.   
This way one has the Penner matrix model \eqref{eq:penner} 
where $N$ is the size of the matrix  determined by the neutrality condition 
\eqref{eq:screeningcondition}.

We now see how the stress tensor insertion in the conformal block
\begin{eqnarray}
\frac{\langle T(z) \mathcal{V}_\infty \mathcal{V}_{0} \cdots \mathcal{V}_n\rangle}{\langle \mathcal{V}_\infty \mathcal{V}_0\cdots \mathcal{V}_n\rangle}
\end{eqnarray}
can be calculated in the Penner type matrix model, where $\langle \mathcal{V}_\infty \mathcal{V}_{0} \cdots \mathcal{V}_n\rangle$ denotes the conformal block associated with the correlator \eqref{eq:correlator}. For that, we first recall that the loop equation of the matrix model is obtained by changing variables as $\lambda_I \to \lambda_I + \frac{\epsilon}{\lambda_I-z}$ and collecting terms of $\mathcal{O}(\epsilon)$ in the partition function \eqref{eq:penner}:
\begin{eqnarray}
0 &=& -\sum_{I,J=1}^N \left\langle \frac{\beta}{(\lambda_I-z)(\lambda_J-z)}\right\rangle - \sum_{I=1}^N \left\langle \frac{1-\beta}{(\lambda_I-z)^2}\right\rangle
\nonumber \\[2mm]
&&\quad  - \frac{\sqrt{\beta}}{g}V'(z)\sum_{I=1}^N\left\langle \frac{1}{z-\lambda_I}\right\rangle 
 + \frac{\sqrt{\beta}}{g}\sum_{I=1}^N\left\langle\frac{V'(z)-V'(\lambda_I)}{z-\lambda_I}\right\rangle.
\label{eq:pre_loop}
\end{eqnarray}
The first and second terms come from the variations of the measure and the Vandermonde determinant while the third and fourth terms come from the variation of the potential. When we define
\begin{eqnarray}
 W(z_1,\cdots,z_s) &=& \beta  \left(\frac{g} {\sqrt{\beta}}\right)^{2-s}\left\langle \sum_{I_1}\frac{1}{z_1-\lambda_{I_1}}\cdots \sum_{I_s}\frac{1}{z_s-\lambda_{I_s}} \right\rangle_{\!\!\rm conn},
\label{eq:Ws}
\\
 f(z) &=& 4g\sqrt{\beta}\sum_{I=1}^N\left\langle \frac{V'(z)-V'(\lambda_I)}{z-\lambda_I}\right\rangle,
\label{eq:f}
\end{eqnarray}
the equation \eqref{eq:pre_loop} can be written as
\be
0=  g^2  W(z,z) +W(z)^2  + V'(z)W(z) +g\left(\sqrt{\beta} -\frac{1}{\sqrt{\beta}}\right)W'(z)   - \frac{f(z)}{4} ,
\label{eq:loop-0}
\ee
which is called the loop equation.
For the Penner type model, the explicit form of the potential \eqref{eq:pot0} implies that
\be
 f(z) = 4 g^2 \sum_{k=0}^n \frac {d_k} {z-z_k},\qquad d_k = \frac{\partial}{\partial z_k} \log Z_{\beta-\rm Penner}.
\ee

Now, let us rewrite the loop equation \eqref{eq:loop-0} as 
\be 
  V'^2 + f + \hbar Q V''
=
x(z)^2  + \hbar Q x'(z)-\hbar^2  W(z,z) \,,
\label{eq:loop-1}
\ee
where $x(z)\equiv 2W(z) + V'(z)$ and $Q= b+ 1/b$.
The left hand side of the equation is illuminating if one writes it as
\be
\varphi(z) \;\equiv\;  -\frac{1}{\hbar^2}(V'^2 + f + \hbar QV'')
\;=\;  \sum_k \frac{\Delta_k}{(z-z_k)^2} 
+ \frac1{z-z_k} \frac{\partial}{\partial z_k} \log Z_{\rm eff},
\label{eq:LHS}
\ee
where $ Z_{\rm eff} \equiv
Z_{\beta-\rm Penner}  \prod_{0 \le a <b \le n} (z_a - z_b)^{-2 \alpha_a \alpha_b} $. Then, equation \eqref{eq:regular_primary} and $Z_{\rm eff} = \langle \mathcal{V}_\infty \mathcal{V}_0\cdots \mathcal{V}_n\rangle$ imply
\begin{eqnarray}
\varphi(z) = \frac{\langle T(z) \mathcal{V}_\infty \mathcal{V}_0\cdots \mathcal{V}_n\rangle}{\langle \mathcal{V}_\infty \mathcal{V}_0\cdots \mathcal{V}_n\rangle}.
\label{eq:stress_tensor}
\end{eqnarray}
In general,  $\varphi(z)$ has the form 
\be
\varphi(z) =  \frac{P_{2n}(z)} {\prod_{k=0}^n (z-z_k) }
\ee
where $P_{2n}(z)$ is a polynomial of $z$ with order $2n$.
This consideration allows us to consider the loop equation as the one generating  the Virasoro constraint. The identification \eqref{eq:stress_tensor} is consistent with the AGT relation \cite{AGT:2009} as will be seen in section \ref{sec:gauge_theories}.

\subsection{Colliding limit of the Penner models}
\label{subsec:colliding_Penner}

We now take the limit of $\alpha_k\to \infty,\,z_k\to 0$ in the matrix model \eqref{eq:penner} while keeping
\begin{eqnarray}
\alpha= \sum_{r=0}^n \alpha_r,\qquad
c_k=  \sum_{r=0}^n \alpha_r \sum_{0 \le s_1 < \cdots < s_k \le n\,(s_i\neq r)} \prod_ {i=1}^k  (-z_{s_i})
\end{eqnarray}
finite. In the CFT side, this gives a conformal block with an irregular vertex operator at $z=0$ and a regular vertex operator at $z=\infty$, or equivalently, the inner product $\langle R_1|I^{(n)}\rangle$. In the matrix model side, the same colliding limit gives the matrix model potential
\begin{eqnarray}
\frac{1}{\hbar} V(z) = \alpha\log z - \sum_{k=1}^n \frac{c_k}{kz^k},
\label{eq:pot1}
\end{eqnarray}
and the neutrality condition $\alpha + \alpha_\infty + bN = Q$. We denote by $Z_M$ the partition function of the matrix model, that is, $Z_M \equiv \lim_{z_r\to0,\alpha_r \to\infty}Z_{\rm \beta-{\rm Penner}}$. We call this type of matrix model ``$D_{2n}$ type'' for a reason which we will see in section \ref{sec:gauge_theories}. This matrix model gives
\be
f(z) =
- \hbar^2  \sum_{k=0}^{n-1} \frac {v_k (\log Z_M )} {z^{2+k}} 
\label{eq:V-f}
\ee 
where  $v_k = \sum_{a=1}^n a c_{a+k} \frac{\partial}{\partial c_a} $ 
using the notation  $c_\ell=0$ if $\ell \ge n+1$.
The term proportional to $1/z$ vanishes due to the identity $\langle \sum_I V'(\lambda_I) \rangle =0$.

Since we have taken the same colliding limit in the CFT side and in the matrix model side, we now identify
\begin{eqnarray}
Z_M = \langle R_1|I^{(n)}\rangle,
\label{eq:identification0}
\end{eqnarray}
where the right hand side is the inner product of an irregular vector $|I^{(n)}\rangle$ of rank $n$ and a regular vector $|R_1\rangle$. Although we have identified the regular conformal block with the rescaled partition function $Z_{\rm eff} = Z_{\beta-{\rm Penner}}\prod_{0\leq a<b\leq n}(z_a-z_b)^{-2\alpha_a\alpha_b}$, we here identify the irregular conformal block with $Z_M$ itself. This difference comes from the definition $|I^{(n)}\rangle \equiv \lim_{z_r\to 0,\alpha_r\to\infty}\left(\prod_{0\leq r<s\leq n}(z_r-z_s)^{2\alpha_r\alpha_s}|R_n\rangle\right)$ of the irregular vector.

It is easy to note that for the potential \eqref{eq:pot1}
\be
V'^2 + \hbar QV''  = - \hbar^2 \sum_{k=0}^{2n} \frac {\Lambda_k}{z^{k+2}}
\ee 
where
 $\Lambda_k = (k+1) Q c_k - \sum _{\ell=0}^{k} c_\ell c_{k-\ell}$ with $\Lambda_0 = \Delta_\alpha = \alpha(Q-\alpha)$.\footnote{Recall here that $c_0 \equiv \alpha$.}
At the colliding limit, we obtain
\be
\varphi (z) \equiv - \frac1{\hbar^2} \left(V'^2 + f + \hbar Q V'' \right)
= \sum_{k=0}^{2n} \frac {\Lambda_k}{z^{k+2}} 
+ \sum_{k=0}^{n-1} \frac {v_k (\log Z_M )} {z^{k+2}}  \,.
\label{phi-n}
\ee
Since $\varphi(z)$ is identified with the stress tensor insertion into the conformal block, this shows that the non-vanishing expectation values of the Virasoro generators are read off as
\be
\frac{ \langle R_1| L_k | I^{(n)} \rangle } { \langle R_1| I^{(n)} \rangle } 
 =  \left\{ 
\begin{array} {l}
 \Lambda_k \quad{\rm when }~ n\le k \le 2n \\ 
 \Lambda_k +v_k (\log Z_M )  \quad{\rm when }~ 0\le k \le n-1\\ 
\end{array}
\right..
\ee
This is in perfect agreement with \eqref{eq:T_irregular} and \eqref{eq:op-CL}, and supports our identification \eqref{eq:identification0}.
Thus, at the colliding limit  the Penner model realizes the simultaneous eigenstate of the Virasoro generators $L_k$ with $n\le k \le 2n$.  On the other hand, $L_k$ with $ k <n$ is represented in terms of differential operator $v_k$. Thus, we will call the colliding limit of the Penner model as the irregular matrix model.

\subsection{Matrix models for general irregular conformal blocks}

The matrix model we have obtained above gives a Liouville conformal block with an irregular vertex operator at $z=0$ and/or a regular vertex operator at $z=\infty$, or equivalently, an inner product $\langle R_1|I^{(n)}\rangle$. The irregular vector $|I^{(n)}\rangle$ is characterized by $\alpha,c_1,\cdots,c_n$ and $v_k(\log Z_M)$ for $k=0,\cdots, n-1$, while the regular vector $|R_1\rangle$ is specified by $\alpha_\infty$. Although the parameter $\alpha_\infty$ does not appear in the matrix model potential \eqref{eq:pot1}, it is encoded in the matrix size $N$ through the condition $\alpha + \alpha_\infty +bN = Q$. In particular, if $\alpha_\infty=0$ then we have no vertex operator insertion at $z=\infty$. As will be seen in section \ref{sec:gauge_theories}, the matrix models for the potential \eqref{eq:pot1} are related to the Argyres-Douglas theories of $A_{2n-3}$ and $D_{2n}$ types.

It is worth noting that the simultaneous eigenvalues of $L_{2n},\cdots, L_{n}$ for the irregular vector $|I^{(n)}\rangle$ are now encoded in the parameters in the matrix model potential $V(z)$. On the other hand, the actions of $L_{n-1},\cdots,L_{0}$ on $|I^{(n)}\rangle$ are specified by $v_k\log (\log Z_{M})$ in the matrix model side. From \eqref{eq:V-f}, we find that $v_k(\log Z_{M})$ are determined when we fix $f(z)$. Since fixing $f(z)$ is equivalent to fixing the filling fractions of the matrix model, the distributions of $N$ eigenvalues specify the actions of $L_{n-1},\cdots,L_{0}$ on the irregular vector.

By generalizing the argument in subsection \ref{subsec:colliding_Penner}, we can now construct matrix models for general Liouville conformal blocks with many regular and irregular vertex operators inserted. In fact, when we consider some additional regular/irregular vertex operators at $z=z_k$ for $k=1,\cdots,r$, the matrix model potential is now given by
\begin{eqnarray}
\frac{1}{\hbar}V(z) = \left(\alpha \log z - \sum_{j=1}^n\frac{c_j}{jz^j}\right) + \sum_{k=1}^{r}\left( \alpha^{(k)}\log(z-z_k) - \sum_{j=1}^{n_k}\frac{c_{j}^{(k)}}{j(z - z_k)^j}\right),
\label{eq:general1}
\end{eqnarray}
with a neutrality condition $\alpha + \sum_{k=1}^{r}\alpha^{(k)} + \alpha_\infty + bN = Q$.
The partition function of this matrix model reproduces the Liouville conformal block with $(r+2)$ vertex operators inserted at $z=0,\infty$ and $z=z_k$. The vertex operator at $z=z_k$ is regular if $c_j^{(k)}=0$ for all $j$, while it is irregular if some $c_j^{(k)}\neq 0$.

It is also possible to make the vertex operator at $z=\infty$ irregular. One way to do this is to consider a colliding limit of some regular vertex operators into $z=\infty$. However, in the matrix model side, we can easily see the effect of an irregular vertex operator at $z=\infty$, just by changing the variables as $\lambda_I\to 1/\lambda_I$ in the matrix model with \eqref{eq:pot1}.
Under this transformation, the integrand of the matrix model integral
\begin{eqnarray}
\int \left[\prod_{I=1}^Nd\lambda_I\right] \Delta_N^{2\beta}\exp\left(\frac{\sqrt{\beta}}{g}\sum_IV(\lambda_I)\right)
\label{eq:matrix_integral}
\end{eqnarray}
gives some changes.
By exponentiating all the changes coming from the integration measures and the Vandermonde determinant, we obtain the same form of the matrix integral with a different potential
\begin{eqnarray}
\frac{1}{\hbar}V(z) = \alpha_\infty \log z - \sum_{k=1}^{n}\frac{c_k z^k}{k},
\end{eqnarray}
where the logarithmic term expresses the regular vertex operator now at $z=0$ and the other terms (together with $\alpha_0$) characterizes the irregular vertex operator at $z=\infty$. Thus, in general, adding some polynomials of $z$ to the potential leads to an irregular vertex operator at $z=\infty$. For example, the matrix model for a Liouville conformal block with irregular vertex operators of rank $m$ at $z=0$ and rank $n$ at $z=\infty$ is given by
\begin{eqnarray}
\frac{1}{\hbar}V(z) = \left(\alpha^{(0)}\log z - \sum_{j=1}^{m}\frac{c_j^{(0)}}{jz^j}\right) - \sum_{k=1}^{n}\frac{c_k^{(\infty)}z^k}{k},
\label{eq:general2}
\end{eqnarray}
with a neutrality condition $\alpha^{(0)} + \alpha^{(\infty)} + bN =Q$. Here $\alpha^{(0)},c_k^{(0)}$ characterizes the irregular singularity at origin, while $\alpha^{(\infty)},\,c_k^{(\infty)}$ characterizes the one at infinity. Of course, we can further add some regular/irregular vertex operators at $z=z_k$, just by adding the second bracket of \eqref{eq:general1} and modify the neutrality condition.

Hence, {\it if the matrix model potential $V(z)$ is written as a sum of logarithmic and/or rational functions of $z$, the matrix model integral \eqref{eq:matrix_integral} gives a Liouville conformal block with regular and/or irregular vertex operators inserted.} Note here that, among the large class of matrix models we have constructed here, a matrix model for two irregular states of rank 1 as well as a model for two regular states and one irregular state of rank 1 were already obtained in \cite{Eguchi:2009gf}. 

As will be seen in section \ref{sec:gauge_theories}, the matrix models for \eqref{eq:general1} and \eqref{eq:general2} are related to $d=4,\mathcal{N}=2$ gauge theories which is not necessarily conformal. In particular, the potential \eqref{eq:general2} is associated with an asymptotically free theory denoted by $\widehat{A}_{2m,2n}$ theory in the notation of \cite{Bonelli:2011aa}.

\section{Irregular partition function}
\label{sec:irregular_partition}

The matrix models we have constructed above can be analyzed by the usual loop equation method with the genus expansion of $\beta$-ensembles \cite{CEM:2009, C:2010, CEM:2010}. As an example, in subsections \ref{subsec:D_2n} and \ref{subsec:D_4}, we illustrate how to calculate the partition function of the $D_{2n}$-type matrix models with potential \eqref{eq:pot1}, order by order in the genus expansion. In particular, we explicitly evaluate the first three expansion coefficients of $\log Z_{M}$ for the $D_4$ matrix model. The result is in perfect agreement with the proposed ansatz \eqref{eq:ansatz} in the Liouville theory side. In subsection \ref{subsec:expansion}, we consider more general matrix models and argue that the small $c_2$ limit of the matrix models is consistent with \eqref{eq:ansatz} at the leading order of the $c_2$-expansion.

\subsection{Property of the $D_{2n}$ type matrix model}
\label{subsec:D_2n}

In this and next subsections, we concentrate on the $D_{2n}$ matrix model with the potential \eqref{eq:pot1}, which gives the inner product of a regular and an irregular conformal block  $\langle R_1|I^{(n)}\rangle$. We here denote by $Z_{M}^{(n)}$ the partition function of the $D_{2n}$ matrix model.

One can evaluate the  partition function using the loop equation
\begin{align}
\frac{f(z)}{4} = \frac{\hbar^2}{4}  W(z,z) +W(z)^2  + V'(z)W(z) + \frac{\hbar Q }2 W'(z),
\label{eq:loop2}
\end{align}
where $f(z)/4$ is now written as
\begin{align}
\frac{f(z)}{4} = - \left(  \frac{\hbar}2 \right)^2~ 
 \sum_{k=0}^{n-1} \frac {v_k (\log Z_M^{(n)} )} {z^{2+k}},\qquad v_k = \sum_{a=1}^{n}ac_{a+k}\frac{\partial}{\partial c_a}.
\end{align}
Recall here that we set $c_k = 0$ for $k>n$. We can regard the loop equation \eqref{eq:loop2} as a series of differential equations for $Z_{M}$. To see this, we expand the quantities around $z=\infty$ as
\begin{align}
 V' (z) &= \hbar  \sum_{a=0}^n  \frac{ c_a }  {z^{a+1}},\qquad W(z) = \frac{\hbar b}2 \sum_{\ell\ge 0} \frac{ \langle \sum_{I=1}^N(\lambda_I)^\ell \rangle}{z^{\ell+1}},
\\
g^2 W(z,z) &= \left(  \frac{\hbar b}2\right)^2  \sum_{\ell , m\ge 1} \frac{ \langle
 \sum_{I=1}^N(\lambda_I)^\ell\sum_{J=1}^N(\lambda_J)^m \rangle_{\rm conn}}{z^{\ell+m+2}}.
\end{align} 
By collecting terms at each order of $z$ in the loop equation \eqref{eq:loop2}, we obtain a system of differential equations.

For example, in the case of $n=1$ where the potential $V(z)$ has two coupling constants $\alpha$ and $c_1$, the loop equation gives us a single differential equation
\be
-v_0 (\log Z_{M}^{(1)} )= A_2  
\label{eq:n=1}
\ee
where $v_0 = c_1\frac{\partial}{\partial c_1}$ and $A_2 = (b^2 N^2 + 2 b \alpha N - bQN) $. 
This differential equation solves the partition function as 
\be
\log Z_{M}^{(1)} = -A_2\log c_1 + {\rm constant}.
\ee
where constant term is independent of $\alpha,c_1,g,b$ and $N$.

In the case $n=2$, the potential $V(z)$ includes three couplings $\alpha,c_1$ and $c_2$, and the loop equation gives two differential equations
\begin{align}
-v_0 (\log Z_{M}^{(2)} ) &= A_2 
\label{eq:v0} \\
-v_1 (\log Z_{M}^{(2)})&= 2 b N c_1 + A_3~ \langle \sum_{I=1}^N\lambda_I \rangle,
\label{eq:n=2}
\end{align}
where $v_0 = c_1\frac{\partial}{\partial c_1} + 2c_2 \frac{\partial}{\partial c_2},\,v_1 = c_2\frac{\partial}{\partial c_1},\, A_3= 2 b^2 N + 2 b  \alpha  - 2b Q$, and $A_2$ is the same quantity as before.
To find $ Z_{M}^{(2)}$, one first notes that there is a homogeneous solution 
to $v_0 = c_1\frac{\partial}{\partial c_1} + 2c_2 \frac{\partial}{\partial c_2}$ since  $v_0 ( H(t))  =0$ for any function of $t \equiv c_2/c_1^2 $.  
Thus, it is convenient to consider  $\log Z_{M}^{(2)}$ as a function of  $t, c_2$ and solve \eqref{eq:v0} as
\be
 \log Z_{M}^{(2)} = -\frac{A_2}2  \log c_2 + H(t) \,. 
\ee
Applying $v_1=c_2\frac{\partial}{\partial c_1}$ to  $ \log Z_{M}^{(2)} $ one has 
\be
v_1  (\log Z_{M}^{(2)}) =  - 2 c_1 t^2 \frac{\partial \log Z_{M}^{(2)} } {\partial t} 
= -2c_1 t^2 H'(t)  \,.
\ee
Then, it follows from equation \eqref{eq:n=2} that
\be
H'(t) = \frac1{2 t^2} \left(A_3~ 
\frac{ \langle \sum_{I=1}^N \lambda_I \rangle}{c_1}  + 2Nb \right) \,. 
\ee
This shows that $\langle \sum_{I=1}^N \lambda_I \rangle/c_1$ is the function of $t$ only.  
Integrating over $t$ one has the solution 
\be 
H(t) =  -\frac {Nb}t  + \frac{A_3}2 \int dt ~\frac {  G_1(t) } {t^2} 
\ee
where $ G_1 (t) \equiv  \langle \sum_{I=1}^{N} \lambda_I \rangle/c_1  $.  Thus, the partition function is given in terms $G_1(t)$. 

For $n\ge 3$  one finds $n$ differential equations for $v_k (\log Z_M)$ for $0\le k \le n-1$
involving  $\langle( {\rm Tr} M^a )  \rangle $ as well as   $\langle( {\rm Tr} M^a ) ~( {\rm Tr} M^b)  \rangle_{\rm conn} $. Among them, we always have
\begin{eqnarray}
-v_0(\log Z_M^{(n)}) = A_2.
\label{eq:general_v0}
\end{eqnarray}
Since any functions of the ratios 
\begin{eqnarray}
\frac{c_2}{c_1^2},\; \frac{c_3}{c_1^3},\; \cdots,\;\frac{c_n}{c_1^n}
\end{eqnarray}
are in the kernel of $v_0$, the differential equation \eqref{eq:general_v0} implies 
\begin{eqnarray}
\log Z_M^{(n)} = -\frac{A_2}{n}\log c_n + H\left(\frac{c_2}{c_1^2},\cdots,\frac{c_n}{c_1^n}\right).
\end{eqnarray}
The function $H$ is determined by other differential equations.

\subsection{Details on $D_4$ type}
\label{subsec:D_4}

In this subsection, we study the first non-trivial example of the $n=2$ case in detail. The potential $V(z)$ and the quantity $f(z)$ are now given by
\begin{eqnarray}
\frac{1}{\hbar}V(z) = \alpha\log z - \frac{c_1}{z} - \frac{c_2}{2z^2},\qquad f(z) = \frac{d_0}{z^2} + \frac{d_1}{z^3},
\label{eq:pot3}
\end{eqnarray}
where we set $d_0 \equiv -\hbar^2 v_0(\log Z_M^{(2)})$ and $d_1\equiv - \hbar^2 v_1(\log Z_M^{(2)})$.
As seen above, the partition function $Z_M^{(2)}$ is given by 
\be
 \log Z_{M}^{(2)} (x, t) = -\frac{A_2}2  \log c_2  +H(t),\qquad H(t) = \int^t \frac{d_{1}}{2\hbar^2 c_1}\frac{dt'}{t'^2}
\label{eq:Z-H}
\ee 
Therefore, when we obtain $d_1/2c_1$ as a function of $t=c_2/c_1^2$, we can evaluate the partition function $Z_{M}^{(2)}$. In fact, the quantity $d_1$ is fixed by {\it the filling fraction.} Since the $D_4$ matrix model has two cuts, there is a single independent filling fraction.

\subsubsection*{$\mathcal{O}(\hbar^{-2})$ contribution}

To be more specific, let us consider the large $N$ expansion of our matrix model while keeping $\hbar N = \mathcal{O}(1)$, and evaluate $\log Z_{M}^{(2)}$ order by order in $\hbar$. In order to use the usual genus expansion method, we set $\alpha,c_1,c_2 = \mathcal{O}(N)$ so that $V(z)$ is of $\mathcal{O}(1)$. In other words, we rescale the parameters as 
\begin{eqnarray}
\alpha = \frac{\hat{\alpha}}{\hbar},\quad c_1 = \frac{\hat{c}_1}{\hbar},\quad c_2 = \frac{\hat{c}_2}{\hbar},
\end{eqnarray}
and treat $\hat{\alpha},\hat{c}_1,\hat{c}_2$ as of $\mathcal{O}(1)$. In this setup, the resolvents and the partition function have the following expansions
\begin{align}
W(z) = \sum_{n=0}^\infty W_n(z)\hbar^n,\quad W(z,z) = \sum_{n=0}^\infty W_n(z,z)\hbar^n,\quad\log Z_{M}^{(2)} = \sum_{n=0}^\infty F_n\, \hbar^{n-2}.
\end{align}
We can also expand $d_0$ and $d_1$ as
\begin{eqnarray}
d_0 = d_{0;0} + \hbar d_{0;1},\qquad d_1 = \sum_{n=0}^\infty d_{1;n}\hbar^{n},
\end{eqnarray}
with $d_{0;0} = b\hbar N(b\hbar N + 2\alpha),\; d_{0;1} = -bQ\hbar N$.
Then, the loop equation implies that at the leading order
\begin{eqnarray}
&&\qquad\qquad \qquad\qquad 2W_0(z) +  V'(z) = \frac{\sqrt{\mathcal{P}_4(z)}}{z^3},
\label{eq:W0}
\\[2mm]
&& \mathcal{P}_4(z) = (d_{0;0} + \hat{\alpha}^2)z^4 + (d_{1;0} + 2\hat{\alpha} \hat{c}_1)z^3 + (\hat{c}_1^2 + 2\hat{\alpha}\hat{c}_2)z^2 + 2\hat{c}_1\hat{c}_2z + \hat{c}_2^2.
\end{eqnarray}
The polynomial $\mathcal{P}_4(z)$ has four roots, which implies that $W_0(z)$ has two branch cuts on $z$-plane. Note here that two of the four roots of $\mathcal{P}_4(z)$ is proportional to $c_2$ while the others are not. Let us denote a branch cut between the first two roots by $\Gamma_2$, and a cut between the other two roots by $\Gamma_1$. Then, in the limit of $c_2\to 0$, the branch cut $\Gamma_2$ shrinks into a point while $\Gamma_1$ still has a finite width in $z$-plane.
Then, we set the filling fraction condition as
\be
 \frac1{2 \pi i} \oint_{\mathcal{A}_1} W(z) dz  = \frac{\hbar b}{2} N_{1},
\label{eq:constraint}
\ee
where $\mathcal{A}_1$ is a cycle encircling the branch cut $\Gamma_1$. Since the total number $N$ of eigenvalues are fixed, we have a constraint $N_{1} + N_{2} = N$ for
\begin{eqnarray}
\frac{1}{2\pi i }\oint_{\mathcal{A}_2}W(z) dz = \frac{\hbar b}{2} N_2,
\end{eqnarray}
where $\mathcal{A}_2$ is a cycle encircling $\Gamma_2$. Below, we see how the filling fraction condition \eqref{eq:constraint} fixes the $\mathcal{O}(\hbar^{-2})$ contribution to the function $H(t)$.

We first assume $N_1$ and $N_2$ are of $\mathcal{O}(N)$, which implies that the right hand side of \eqref{eq:constraint} is of $\mathcal{O}(1)$. Then, the leading contribution to \eqref{eq:constraint} gives
\begin{eqnarray}
\hbar b N_1 = \frac{1}{\pi i}\oint_{\mathcal{A}_1}W_0(z) dz = \frac{1}{2\pi i}\oint_{\mathcal{A}_1}\frac{\sqrt{\mathcal{P}_4(z)}}{z^3}dz.
\label{eq:constraint2}
\end{eqnarray}
Since $\mathcal{A}_1$ encircles the cut $\Gamma_1$ which does not shrink in the limit $c_2\to 0$, we can expand the integral around $\hat{c}_2 = 0$ as
\begin{eqnarray}
\oint_{\mathcal{A}_1}\!\! dz\frac{\sqrt{\mathcal{P}_4(z)}}{z^3} &=& \oint_{\mathcal{A}_1}\!\! dz\frac{\sqrt{\mathcal{Q}_4(z)}}{z^3}\times 
\nonumber\\
&&\qquad\left(1 + \frac{2\hat{\alpha}\hat{c}_2 z^2 + 2\hat{c}_1\hat{c}_2 z + \hat{c}_2^2}{2\mathcal{Q}_4(z)}- \frac{(2\hat{\alpha}\hat{c}_2 z^2 + 2\hat{c}_1\hat{c}_2 z + \hat{c}_2^2)^2}{8\mathcal{Q}_4(z)^2} - \cdots \right),
\nonumber\\[2mm]
&=& \oint_{\mathcal{A}_1}\!\!\frac{\sqrt{\mathcal{Q}_4(z)}}{z^3}dz + \hat{c}_2 \oint_{\mathcal{A}_1}\frac{\hat{\alpha}z + \hat{c}_1}{z^2\sqrt{\mathcal{Q}_4(z)}}dz + \mathcal{O}(\hat{c}_2^2)
\label{eq:integral1}
\end{eqnarray}
where
\begin{eqnarray}
\mathcal{Q}_4(z) \equiv (d_{0;0} + \hat{\alpha}^2)z^4 + (d_{1;0}+ 2\hat{\alpha}\hat{c}_1)z^3 + \hat{c}_1^2z^2
\end{eqnarray}
is independent of $c_2$. Since $\sqrt{\mathcal{Q}_4(z)}$ has only one branch cut $\Gamma_1$, we can evaluate the right-hand side of \eqref{eq:integral1} just by taking residues at $z=0$ and $z=\infty$, that is,
\begin{eqnarray}
&& \frac{1}{2\pi i}\oint_{\mathcal{A}_1}\!\!\frac{\sqrt{\mathcal{P}_4(z)}}{z^3}dz = \left(b \hbar N -\frac{d_{1;0}}{2\hat{c}_1}\right) + \frac{ \left(4 \hat{c}_1^2 d_{0;0}-3d_{1;0}^2-8\hat{\alpha} \hat{c}_1d_{1;0} \right)}{8 \hat{c}_1^4}\hat{c}_2 +\mathcal{O}(\hat{c}_2^2)
\nonumber\\
\end{eqnarray}
From this and the filling fraction condition \eqref{eq:constraint2}, we find 
\begin{eqnarray}
\frac{d_{1;0}}{2\hat{c}_1} &=& b\hbar N_2 +  \frac{1}{2}\Big\{b\hbar N_1(b\hbar N_1 +2\hat{\alpha}) - 2b\hbar{N}_2(b\hbar N_2 + \hat{\alpha}) + 2b^2\hbar^2N_1N_2\Big\}\hat{t} + \mathcal{O}(\hat{t}^2),
\nonumber\\
\end{eqnarray}
where $\hat{t}=\hat{c}_2/\hat{c}_1^2$. Then, \eqref{eq:Z-H} implies that the $\mathcal{O}(\hbar^{-2})$ contribution to $\log Z_{M}^{(2)}$ is evaluated as
\begin{eqnarray}
\hbar^{-2}F_0 &=& -\frac{bN_2}{t} - \frac{bN(bN+2\alpha)}{2}\log c_2
\nonumber\\
&&+ \frac{1}{2}\left\{ bN_1(bN_1+2\alpha) - 2bN_2(bN_2 + \alpha) + 2b^2 N_1N_2\right\}\log t
+\mathcal{O}(t),
\label{eq:genus0}
\end{eqnarray}
up to constant.

\subsubsection*{$\mathcal{O}(\hbar^{-1})$ contribution}

Let us evaluate the next-to-leading contribution to $\log Z_{M}^{(2)}$. By collecting terms of $\mathcal{O}(\hbar)$ in the loop equation, we obtain
\begin{eqnarray}
 W_1(z) &=& \frac{z^3}{\sqrt{\mathcal{P}_4(z)}}\left(-\frac{Q}{2}W_0'(z) + \frac{d_{0,1}}{4z^2} + \frac{d_{1,1}}{4z^3}\right)
\nonumber\\
&=& \frac{1}{4\sqrt{\mathcal{P}_4(z)}}\left(Qz^3V''(z) +z d_{0,1} + d_{1,1}\right) - \frac{Q}{8}(\log \mathcal{P}_4(z))'+ \frac{3Q}{4z}.
\label{eq:W1}
\end{eqnarray}
Then its period is evaluated as
\begin{eqnarray}
\frac{1}{2\pi i}\oint_{\mathcal{A}_1}\!\!W_1(z)dz &=& \frac{1}{2\pi i}\oint_{\mathcal{A}_1}\!\!\frac{dz}{4\sqrt{\mathcal{P}_4(z)}}\left(Qz^3V''(z) +z d_{0;1} + d_{1;1}\right) - \frac{Q}{4}.
\label{eq:period1}
\end{eqnarray}
Here, the integral in the right-hand side can be expanded in powers of $c_2$ as
\begin{eqnarray}
&&\!\!\!\!\!\!\!\!\!\!\!\! \oint_{\mathcal{A}_1}\frac{dz}{4\sqrt{\mathcal{P}_4(z)}}\left(Qz^3V''(z) +z d_{0;1} + d_{1;1}\right)
\\
&=& \oint_{\mathcal{A}_1}\frac{(d_{0;1}-Q\hat{\alpha})z + (d_{1;1}-2Q\hat{c}_1)}{4\sqrt{\mathcal{Q}_4(z)}}dz + \hat{c}_2\oint_{\mathcal{A}_1}\Bigg(-\frac{3Q}{4z\sqrt{\mathcal{Q}_4(z)}}
\nonumber\\
&&\qquad  -\frac{\hat{\alpha}(d_{0;1}-Q\hat{\alpha})z^3 + (\hat{\alpha}d_{1;1} + \hat{c}_1 d_{0;1} -3Q\hat{\alpha}\hat{c}_1)z^2 + \hat{c}_1(d_{1;1}-2Q\hat{c}_1)z}{4\mathcal{Q}_4(z)^{3/2}}\Bigg)dz
\nonumber\\
&&+ \mathcal{O}(\hat{c}_2^2)
\end{eqnarray}
The right-hand side of this equation is evaluated by picking up residues at $z=0$ and $z=\infty$. Then, \eqref{eq:period1} is written as
\begin{eqnarray}
&&\!\!\!\!\!\!\!\! \frac{1}{2\pi i}\oint_{\mathcal{A}_1}W_1(z)dz = -\frac{d_{1;1}}{4\hat{c}_1}  + \frac{2\hat{c}_1^2 d_{0;1} - 3d_{1;0}d_{1;1} + 3Q\hat{c}_1d_{1;0} - 4\hat{c}_1\hat{\alpha}d_{1;1}}{8\hat{c}_1^4}\hat{c}_2 + \mathcal{O}(\hat{c}_2^2)
\nonumber\\
\end{eqnarray}

Now, recall that we have the filling fraction condition \eqref{eq:constraint}. The $\mathcal{O}(\hbar^{-1})$ contribution to \eqref{eq:constraint} gives
\begin{eqnarray}
\frac{1}{2\pi i}\oint_{\mathcal{A}_1}W_1(z)dz = 0,
\end{eqnarray}
which implies that 
\begin{eqnarray}
\frac{d_{1;1}}{2\hat{c}_1} = \frac{Qb\hbar(2N_2-N_1)}{2}\hat{t}  + \mathcal{O}(\hat{t}^2).
\end{eqnarray}
Here we used $d_{0;1} = -b Q\hbar N$ and $d_{1:0}= 2\hat{c}_1b\hbar N_2+\mathcal{O}(\hat{t})$.
Then, by using \eqref{eq:Z-H}, the $\mathcal{O}(\hbar^{-1})$ contribution to $\log Z_{M}^{(2)}$ is evaluated as
\begin{eqnarray}
\hbar^{-1}F_1 = \frac{QbN}{2}\log c_2 + \frac{Qb(2N_2-N_1)}{2}\log t + \mathcal{O}(t),
\label{eq:genus1/2}
\end{eqnarray}
up to constants.

\subsubsection*{$\mathcal{O}(\hbar^{0})$ contribution}

We now evaluate the $\mathcal{O}(\hbar^{0})$ correction to $\log Z_{M}^{(2)}$. The loop equation \eqref{eq:loop2} at $\mathcal{O}(\hbar^0)$ tells us
\begin{eqnarray}
\frac{d_{1;2}}{4z^3} = W_0(z,z) + 2W_0(z)W_2(z) + W_1(z)^2 + V'(z)W_2(z) + \frac{Q}{2}W_1'(z),
\end{eqnarray}
which implies that
\begin{eqnarray}
W_2(z) = \frac{z^3}{\sqrt{\mathcal{P}_4(z)}}\left(\frac{d_{1;2}}{4z^3} - W_0(z,z) -W_1(z)^2 - \frac{Q}{2}W_1'(z)\right).
\label{eq:W2}
\end{eqnarray}
Therefore, we first need to evaluate $W_0(z,z)$ to calculate $W_2(z)$.

In fact, $W_0(z,z)$ is obtained from an another identity for the resolvents. By changing the variables as
\begin{eqnarray}
\lambda_I \to \lambda_I + \sum_{J=1}^N\frac{\epsilon}{(z_1-\lambda_I)(z_2-\lambda_J)}
\end{eqnarray}
in the matrix integral \eqref{eq:matrix_integral} and collecting terms of $\mathcal{O}(\epsilon)$, we obtain an identity
\begin{eqnarray}
0 &=& -\frac{\hbar^2}{4}W(z_1,z_1,z_2) + (2W(z_1)+V'(z_1))W(z_1,z_2) + \frac{\partial}{\partial z_2}\frac{W(z_1)-W(z_2)}{z_1-z_2}
\nonumber\\
&& -U(z_1,z_2) + \frac{\hbar Q}{2}\frac{\partial }{\partial z_1}W(z_1,z_2),
\end{eqnarray}
where
\begin{eqnarray}
U(z_1,z_2) = \beta\left\langle \sum_{I}\frac{V'(z_1)-V'(\lambda_I)}{z_1-\lambda_I}\sum_{J}\frac{1}{z_2 - \lambda_J}\right\rangle_{\rm connected}.
\end{eqnarray}
From the leading order of this identity, it follows that
\begin{eqnarray}
W_0(z,z) = \frac{U_0(z,z)- W_0''(z)/2}{2W_0(z) + V'(z)}.
\end{eqnarray}
Furthermore, by using another identities
\begin{eqnarray}
\beta\left\langle \sum_I\frac{1}{\lambda_I}\sum_J\frac{1}{z-\lambda_J}\right\rangle_{\rm conn} &=& -\frac{\partial W(z)}{\partial \hat{c}_1}
\\
\beta\left\langle \sum_IV'(\lambda_I)\sum_J\frac{1}{z-\lambda_J}\right\rangle_{\rm conn} &=& W'(z),
\\
\beta \left\langle\sum_I \lambda_IV'(\lambda_I)\sum_J\frac{1}{z-\lambda_J}\right\rangle_{\rm conn} &=& W(z) + zW'(z).
\end{eqnarray}
 $U(z,z)$ can be rewritten as
\begin{eqnarray}
U(z,z) &=& \frac{\hat{c}_2}{z^3}\frac{\partial W(z)}{\partial \hat{c}_1} -  \frac{W(z)}{z^2} - \frac{2W'(z)}{z}.
\end{eqnarray}
Therefore $W_0(z,z)$ can be written as
\begin{eqnarray}
W_0(z,z) &=& \frac{z^3}{\sqrt{\mathcal{P}_4(z)}}\left(\frac{\hat{c}_2}{z^3}\frac{\partial W_0(z)}{\partial\hat{c}_1} - \frac{W_0(z)}{z^2} - \frac{2W_0'(z)}{z} - \frac{W_0''(z)}{2}\right).
\label{eq:W00}
\end{eqnarray}

Then, equations \eqref{eq:W2},\eqref{eq:W1},\eqref{eq:W0} and \eqref{eq:W00} lead to
\begin{eqnarray}
W_2(z) &=& \frac{d_{1;2}}{4\sqrt{\mathcal{P}_4(z)}} -\frac{z(1+\frac{Q^2}{16})}{\sqrt{\mathcal{P}_4(z)}}
\nonumber\\[2mm]
&&+\frac{z \left(-3\hat{c}_2^2-\hat{c}_2 z (3 \hat{c}_1+\hat{\alpha}z)+z^2 \left(\hat{c}_1^2+6\hat{c}_1\hat{\alpha}z +3z\left(d_{1;0}+2z\left(d_{0;0}+\hat{\alpha}^2\right)\right)\right)\right)}{6\mathcal{P}_4(z)^{3/2}}
\nonumber\\
&& + \frac{1}{96\mathcal{P}_4(z)^{3/2}}z \Bigg\{-6 (3 Q\hat{c}_2 + z (-d_{1;1}+2Q\hat{c}_1 -d_{0;1}z+Q\hat{\alpha}z ))^2
\nonumber\\
&&\qquad + 2Q^2 z^2 \bigg(\hat{c}_1^2+2\hat{c}_2\hat{\alpha} +3z(d_{1;0}+2 \hat{c}_1 \hat{\alpha} )+6z^2 \left(d_{0;0}+\alpha^2\right)\bigg)\Bigg\},
\end{eqnarray}
up to total derivative terms which does not contribute to the period.
Then the period $\oint_{\mathcal{A}_1}W_2(z)dz$ has the following $c_2$-expansion:
\begin{eqnarray}
 \frac{1}{2\pi i}\oint_{\mathcal{A}_1}W_2(z)dz &=& \frac{3 b \hbar(4N_2 - (4N_2-N_1)Q^2)}{8}\hat{t}^2 + \Bigg(-\frac{1}{2} - \frac{3b\hbar N_2 + 2\hat{\alpha}}{2}\hat{t}
\nonumber\\
&& - \frac{3(b^2\hbar^2(13N_2^2 - 4N_1N_2 -2N_1^2) + 4b\hbar(4N_2-N_1)\hat{\alpha} + 4\hat{\alpha}^2)}{4}\hat{t}^2\Bigg)\frac{d_{1;2}}{2\hat{c}_1}
\nonumber\\
&& +\mathcal{O}(\hat{t}^3).
\end{eqnarray}
Then the filling fraction condition \eqref{eq:constraint} implies that
\begin{eqnarray}
\frac{d_{1;2}}{2\hat{c}_1} = \frac{3b\hbar(4N_2-(4N_2-N_1)Q^2)}{4}\hat{t}^2 + \mathcal{O}(\hat{t}^3).
\end{eqnarray}
This and \eqref{eq:Z-H} give the $\mathcal{O}(\hbar^{0})$ correction to $\log Z_{M}^{(2)}$ as
\begin{eqnarray}
F_2 = \frac{3b(4N_2 -(4N_2-N_1)Q^2)}{4}t +\mathcal{O}(t^2),
\label{eq:genus1}
\end{eqnarray}
up to constants.

\subsubsection*{Partition function}

From \eqref{eq:genus0},\eqref{eq:genus1/2} and \eqref{eq:genus1}, we find that the partition function $Z_{M}^{(2)}$ of the $D_{4}$ matrix model is written as
\begin{eqnarray}
 Z_M^{(2)} &= 
(c_2)^{- \frac{bN (bN+ 2  \alpha -Q )}2} 
\left( \frac{c_2}{c_1^2}  \right) ^{
     \frac{bN_{1} ( b N_{1}+  2\alpha  - Q )}{2}  -  bN_{2} (b N_{2} + \alpha -Q ) 
+ b^2N_{1} N_{2}
} ~e^{-  \frac {b N_{2} c_1^2  } {c_2 }    + \CO\left(\frac{c_2}{c_1^2}\right)   }.\qquad
\label{eq:partition_genus1}
\end{eqnarray}
in the genus-one approximation of the genus expansion.

Recall that we have identified this partition function with the irregular conformal block $\langle R_1|I^{(2)}\rangle$. Since $|I^{(2)}\rangle$ satisfies \eqref{eq:ansatz} and $|R_1\rangle$ is independent of $c_2$, the partition function \eqref{eq:partition_genus1} should reproduce the small $c_2$ behavior of \eqref{eq:ansatz}.  Let us now check this. In the small $c_2$ limit, the partition function \eqref{eq:partition_genus1} is proportional to
\begin{eqnarray}
c_2^{-bN_2(\frac{3}{2}bN_2 + 2\alpha- \frac{3}{2}Q)}\exp(-\frac{bN_2c_1^2}{c_2}).
\end{eqnarray}
When we identify $-bN_2 = \alpha-\alpha'$, this factor correctly matches \eqref{eq:ansatz}. The identification
\begin{eqnarray}
-bN_2 = -i\sqrt{\beta}N_2 = \alpha - \alpha'
\end{eqnarray}
is reasonable because $-bN_2$ is originally the momentum of an intermediate state in the Liouville (regular) conformal block which should be identified with $\alpha-\alpha'$.

\subsection{Small $c_2$ limit in general}
\label{subsec:expansion}

In this subsection, we consider more general matrix model whose potential is of the form
\begin{eqnarray}
\frac{1}{\hbar}V(z) = \alpha\log z - \frac{c_1}{z} - \frac{c_2}{2z^2} + G(z).
\label{eq:pot2}
\end{eqnarray}
Here $G(z)$ is written as a sum of logarithmic and/or rational functions which are regular at $z=0$. In the Liouville side, the partition function of this matrix model gives a conformal block $\langle G|I^{(2)}(c_1,c_2,\alpha)\rangle$ with an irregular vertex operator of rank $2$ at $z=0$ as well as regular and/or irregular vertex operators inserted away from $z=0$. The vertex operators away from origin are characterized by $G(z)$. We will not explicitly evaluate the partition function of this matrix model. Instead, we here consider the small $c_2$ limit of this matrix model and compare it with the proposed ansatz \eqref{eq:ansatz} of the rank 2 irregular vector $|I^{(2)}\rangle$, including the shift of the momentum $\alpha\to\alpha'$.

The matrix model with potential \eqref{eq:pot2} generally have multiple cuts, among which a single special cut shrinks into a point in the limit of $c_2\to 0$.
When the matrix model has $p$ cuts, the spectral curve is the double cover of a Riemann sphere with $p$ square-root cuts on it. Here, we denote by $N_{i}$ the number of eigenvalues distributed on the $i$-th cut, and fix them with a condition $\sum_{i=1}^{p} N_i = N$.  This implies that, we choose the integration contour $C_i$ of the $N_i$ eigenvalues so that it passes through the $i$-th cut and does not pass through the other cuts. Without loss of generality, we set the $p$-th cut to be the shrinking cut in the limit $c_2\to 0$ and therefore $N_p$ eigenvalues are distributed on the shrinking cut. Then the partition function of the matrix model is written as
\begin{eqnarray}
Z_{\vec{N};N_p} &=& \int_{\vec{C}}\left(\prod_{I=1}^{|\vec{N}|}d\lambda_I\right)\Delta(\lambda)^{2\beta}\;Z_{N_p}(\lambda)\; \exp\left[\frac{\sqrt{\beta}}{g}\sum_{I=1}^{|\vec{N}|}V(\lambda_I)\right],
\label{eq:Z-N0-N}
\end{eqnarray}
where $\vec{C}=(C_1,\cdots,C_{p-1}),\, \vec{N}=(N_1,\cdots,N_{p-1})$ and
\begin{eqnarray}
Z_{N_p}(\lambda) &\equiv& \int_{C_p}\left(\prod_{J=1}^{N_p}d\xi_J\right)\Delta(\xi)^{2\beta} \prod_{I=1}^{|\vec{N}|}\prod_{J=1}^{N_p}(\lambda_I-\xi_J)^{2\beta} \exp\left[\frac{\sqrt{\beta}}{g}\sum_{J=1}^{N_p}V(\xi_J)\right].
\label{eq:reduced_Z}
\end{eqnarray}

 Below, we evaluate the leading contribution in the limit $c_2\to 0$. In the small $c_2$ limit, the $\xi$-integral becomes singular because $N_p$ eigenvalues are distributed on the shrinking cut, while the $\lambda$-integral remains regular because no eigenvalue is distributed on it. In order to study the leading singularity in $Z_{N_p}(\lambda)$ we first rescale $\xi_I$ as $\xi_I\to c_2\xi_I$, which leads to
\begin{eqnarray}
 \frac{1}{\hbar}V(c_2z) = \frac{1}{c_2}\left(-\frac{1}{2z^2} - \frac{c_1}{z}\right) + \alpha\log(c_2 z) + G(c_2z).
\label{eq:c2-rescale}
\end{eqnarray}
Since $G(z)$ is regular at $z=0$, $G(c_2z)$ approaches to a finite constant in the limit of $c_2\to 0$. Therefore, we can omit it in calculating the leading contribution to $Z_{N_p}(\lambda)$. Note here that, since the width of the shrinking cut was of order $\mathcal{O}(c_2)$ before rescaling, it is of $\mathcal{O}(1)$ in $\xi$-plane after the rescaling.\footnote{All the other cuts now shrink into $\xi = \infty$.} Then the leading contribution to $Z_{N_p}(\lambda)$ is now written as
\begin{eqnarray}
 Z_{N_p}(\lambda) = c_2^{(\sqrt{\beta}N_p)^2 + iQ\sqrt{\beta}N_p -2i\alpha \sqrt{\beta}N_p}\prod_{I=1}^{|\vec{N}|}(\lambda_I)^{2\beta N_p}\; \exp \widetilde{F}
\label{eq:xi-matrix}
\end{eqnarray}
where the first factor comes from the logarithmic term in \eqref{eq:c2-rescale} as well as rescaling $d\xi_J$ and $\Delta(\xi)^{2\beta}$. The second factor is the leading contribution from $\prod_{I,J}(\lambda_I-c_2\xi_J)^{2\beta}$.
The remaining contributions are included in $\exp \widetilde{F}$ which is written as (up to a constant prefactor)
\begin{eqnarray}
\exp \widetilde{F} \;\equiv\; \int_{C_p}\big(\prod_{I=1}^{N_p}d\xi_I\big)\Delta(\xi)^{2\beta}  \exp\left[\frac{2\sqrt{\beta}}{ic_2}\sum_{I}\widetilde{V}(\xi)\right]
\label{eq:expF}
\end{eqnarray}
with the reduced potential
\begin{eqnarray}
\widetilde{V}(z) = -\frac{1}{2z^2} - \frac{c_1}{z} + \alpha c_2\log z.
\label{eq:reduced_pot}
\end{eqnarray}
Note here that the term $\alpha c_2 \log z$ gives a finite contribution to $\frac{2\sqrt{\beta}}{ic_2}\sum_I\widetilde{V}(\xi_I)$ even in the limit $c_2\to 0$.

What we need to do next is to evaluate the leading contribution to \eqref{eq:expF}. For that, we use the loop equation for the reduced matrix model integral \eqref{eq:expF}
\begin{eqnarray}
(ic_2/2)^2 \widetilde{W}(z,z) + \widetilde{W}(z)^2 +(c_2/2)Q\widetilde{W}'(z)+ \widetilde{W}(z)\widetilde{V}'(z) - \frac{\widetilde{f}(z)}{4} = 0,
\end{eqnarray}
where we treat $ic_2/2$ as a matrix model coupling constant to define
\begin{eqnarray}
\widetilde{W}(z_1,\cdots,z_s) &\equiv& \beta\left(\frac{ic_2/2}{\sqrt{\beta}}\right)^{2-s}\left\langle \sum_{I_1=1}^{N_p}\frac{1}{z-\xi_{I_1}}\cdots \sum_{I_s}^{N_p}\frac{1}{z-\xi_{I_s}}\right\rangle_{\rm conn.},
\\[2mm]
\widetilde{f}(z) &\equiv& 4(ic_2/2)\sqrt{\beta}\left\langle \sum_{I=1}^{N_p} \frac{\widetilde{V}'(z) - \widetilde{V}'(\xi_I)}{z-\xi_I}\right\rangle.
\end{eqnarray}
Here and in the rest of this subsection, $\langle \mathcal{O}\rangle$ stands for the expectation value of $\mathcal{O}$ with the reduced matrix integral \eqref{eq:expF}. 
From the explicit expression for $\widetilde{V}$ and the identity $\langle\sum_{I=1}^{N_p} \widetilde{V}'(\xi_I)\rangle=0$, we obtain
$\widetilde{f}(z) =  \widetilde{d}_0/z^2 + \widetilde{d}_1/z^3$,
with
\begin{eqnarray}
 \widetilde{d}_0 = -2ic_2\sqrt{\beta}\left\langle \sum_I\left(\frac{1}{\xi_I^2} + \frac{c_1}{\xi_I}\right)\right\rangle,\qquad
 \widetilde{d}_1 = -2ic_2\sqrt{\beta}\left\langle \sum_I \frac{1}{\xi_I}\right\rangle,
\end{eqnarray}
Since the asymptotic behaviors of the resolvents are given by
$\widetilde{W}(z) = (ic_2/2)\sqrt{\beta} N_p/z + (ic_2/2)\sqrt{\beta}\langle\sum_I^{N_p}\xi_I \rangle/z^2 + \mathcal{O}\left(z^{-3}\right),\, \widetilde{W}(z,z) = \mathcal{O}\left(z^{-4}\right)$, the loop equation implies that
\begin{eqnarray}
\widetilde{d}_0 &=& -c_2^2\left(\beta N_p^2 - 2i\sqrt{\beta}\alpha N_p + \sqrt{\beta}iQN_p\right),
\label{eq:D0-D1-0}
\\[2mm]
\widetilde{d}_1 &=& 2c_1c_2i\sqrt{\beta}N_p - c_2^2\left(2\beta N_p - 2\alpha i\sqrt{\beta} + 2\sqrt{\beta}iQ\right)\langle \sum_I\xi_I\rangle.
\label{eq:D0-D1}
\end{eqnarray}
Note that these are exact expressions without any approximation. The equations \eqref{eq:D0-D1-0} and \eqref{eq:D0-D1} imply that all the quantities of the reduced matrix model \eqref{eq:expF}, including $\exp \widetilde{F}$ itself, are completely determined by $\langle \sum_{I=1}^{N_p}\xi_I\rangle$. For example, the free energy $\widetilde{F}$ is obtained by solving
\begin{eqnarray}
c_2\frac{\partial \widetilde{F}}{\partial c_2} =
\frac{\widetilde{d}_0 + c_1\widetilde{d}_1}{2c_2^2},\qquad 
\frac{\partial \widetilde{F}}{\partial c_1} = -\frac{\widetilde{d}_1}{c_2^2}.
\label{eq:dF}
\end{eqnarray}

Especially in the small $c_2$ limit, the eigenvalues are localized at a value that extremizes the potential $\widetilde{V}(\xi_I)$, that is, $\xi_I = -1/c_1$. Note here that the Coulomb repulsion due to $\Delta(\xi)^2$ is subleading and all the $N_p$ eigenvalues take the same value $-1/c_1$ in the limit $c_2\to 0$. The reason for this is that we here keep $N_p$ {\it finite} which is different from the usual 't Hooft expansion of the matrix model.
By substituting $\xi_I=-1/c_1 + \mathcal{O}(c_2)$, we obtain
$\langle \sum_{I=1}^{N_p}\xi_I\rangle = -N_p/c_1 + \mathcal{O}(c_2)$,
and therefore the equation \eqref{eq:D0-D1} implies
\begin{eqnarray}
\widetilde{d}_1 =  2c_1c_2 i\sqrt{\beta} N_p +\frac{(2\beta N_p^2 +2(Q-\alpha)i\sqrt{\beta}N_p)c_2^2}{c_1} + \mathcal{O}(c_2^3).
\end{eqnarray}
Then \eqref{eq:dF} is now written as
\begin{eqnarray}
 c_2\frac{\partial \widetilde{F}}{\partial c_2} &=& \frac{i\sqrt{\beta}N_pc_1^2}{c_2} + \frac{\beta N_p^2}{2} + \frac{iQ\sqrt{\beta}N_p}{2} + \mathcal{O}(c_2),\\[2mm]
\frac{\partial \widetilde{F}}{\partial c_1} &=& -\frac{2i\sqrt{\beta}N_pc_1}{c_2}-\left(2\beta N_p^2 + 2\left(Q-\alpha\right)i\sqrt{\beta}N_p\right)\frac{1}{c_1} + \mathcal{O}(c_2).
\end{eqnarray}
The solution to these differential equations in the limit $c_2\to 0$ is given by
\begin{eqnarray}
\widetilde{F} = -\frac{i\sqrt{\beta}N_p c_1^2}{c_2} -\left(2\beta N_p^2 + 2\left(Q-\alpha\right)i\sqrt{\beta}N_p\right)\log c_1 + \left(\frac{\beta N_p^2}{2}+ \frac{iQ\sqrt{\beta}N_p}{2}\right)\log c_2,
\nonumber\\
\end{eqnarray}
up to constant independent of $c_1$ and $c_2$.
This leading behavior of $\widetilde{F}$ and equations \eqref{eq:Z-N0-N}, \eqref{eq:xi-matrix} imply that the original matrix model behaves in the small $c_2$ limit as
\begin{eqnarray}
Z_{\vec{N};N_p} &\sim& c_2^{-i\sqrt{\beta}N_p(\frac{3}{2}i\sqrt{\beta}N_p +2\alpha - \frac{3}{2}Q)} c_1^{-i\sqrt{\beta}N_p(-2i\sqrt{\beta}N_p-2\alpha +2Q)}\exp\left(-i\sqrt{\beta}N_p\frac{c_1^2}{c_2}\right)
\nonumber
\\[2mm]
&&\quad \times  \int_{\vec{C}}\left(\prod_{I=1}^{|\vec{N}|}d\lambda_I\right)\Delta(\lambda)^{2\beta} \exp\left[\frac{\sqrt{\beta}}{g}\sum_{I=1}^{|\vec{N}|}\widehat{V}(\lambda_I)\right],
\label{eq:small_c2}
\end{eqnarray}
where $\widehat{V}(z)$ is given by
\begin{eqnarray}
 \frac{i}{2g}\widehat{V}(z) = \left(\alpha + i\sqrt{\beta}N_p\right)\log z - \frac{c_1}{z} + G(z).
\label{eq:hat-pot}
\end{eqnarray}
Note here that this potential is independent of $c_2$ and the coefficient of $\log z$ is shifted by $i\sqrt{\beta}N_p$.

The remaining matrix integral \eqref{eq:small_c2} is in fact identified with an irregular conformal block with one irregular singularity of rank {\it one} at $z=0$ and irregular/regular singularities away from $z=0$, the latter of which is characterized by the function $G(z)$. Hence, \eqref{eq:small_c2} implies that in the small $c_2$ limit
\begin{eqnarray}
\langle G| I^{(2)}(c_1,c_2,\alpha)\rangle &\sim&  c_2^{-i\sqrt{\beta}N_p(\frac{3}{2}i\sqrt{\beta}N_p +2\alpha - \frac{3}{2}Q)} c_1^{-i\sqrt{\beta}N_p(-2i\sqrt{\beta}N_p -2\alpha +2Q)}\exp\left(-i\sqrt{\beta}N_p\frac{c_1^2}{c_2}\right)
\nonumber\\
&&\qquad \times \langle G| I^{(1)}(c_1,\alpha +i\sqrt{\beta}N_p)\rangle.
\label{eq:leading_matrix_state}
\end{eqnarray}
When we identify
\begin{eqnarray}
-i\sqrt{\beta}N_p = \alpha-\alpha',
\label{eq:identification}
\end{eqnarray}
we find that \eqref{eq:leading_matrix_state} is in perfect agreement with the leading term of \eqref{eq:ansatz}, including the shift of the momentum $\alpha\to \alpha'$. In fact, the identification \eqref{eq:identification} is the expected one because $i\sqrt{\beta}N_p$ is interpreted as the internal momentum of the regular Liouville conformal block before taking the colliding limit. Since equation \eqref{eq:leading_matrix_state} holds for general $G(z)$ as long as $G(z)$ is regular at $z=0$, it supports the ansatz \eqref{eq:ansatz} proposed in \cite{GT:2012}. We here emphasize that the result of this subsection is valid for all orders of the genus expansion.

\section{Interpretation in gauge theories}
\label{sec:gauge_theories}

It was pointed out \cite{Bonelli:2011aa, GT:2012} that the irregular conformal blocks of the Liouville theory should reproduce the Nekrasov partition function of Argyres-Douglas theories, which is based on the observation that the colliding limits in the gauge theory side give Argyres-Douglas theories and some asymptotically free $\mathcal{N}=2$ theories involving the Argyres-Douglas theories \cite{Gaiotto:2009hg, Xie:2012hs}. Since we have already constructed matrix models for the irregular conformal blocks, we here explore the relation between our matrix models for irregular conformal blocks and the Argyres-Douglas theories in four dimensions.

\subsection{AGT relation}

\begin{figure}
\begin{center}
\includegraphics[width=7cm]{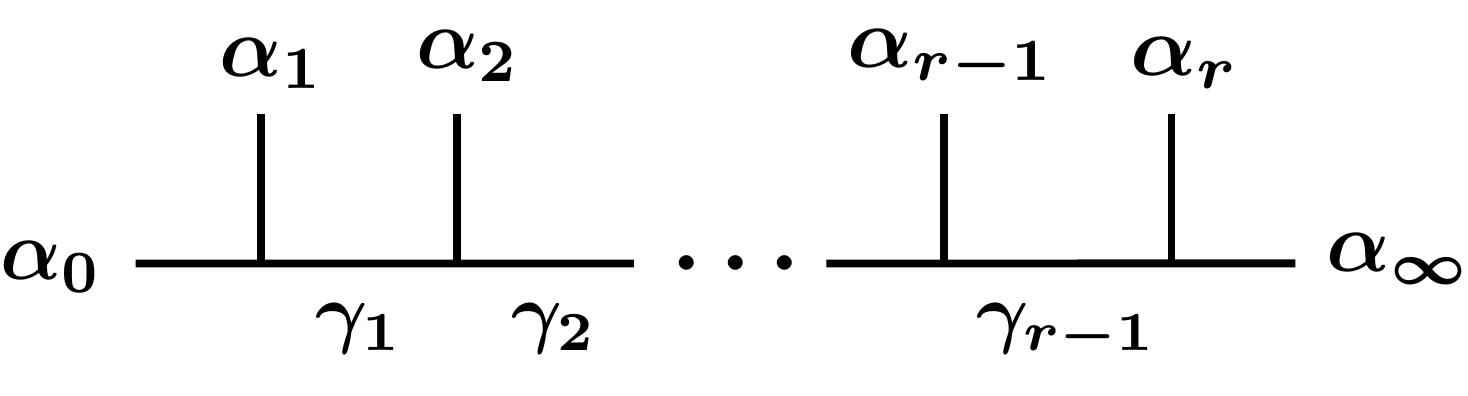}\qquad
\includegraphics[width=7cm]{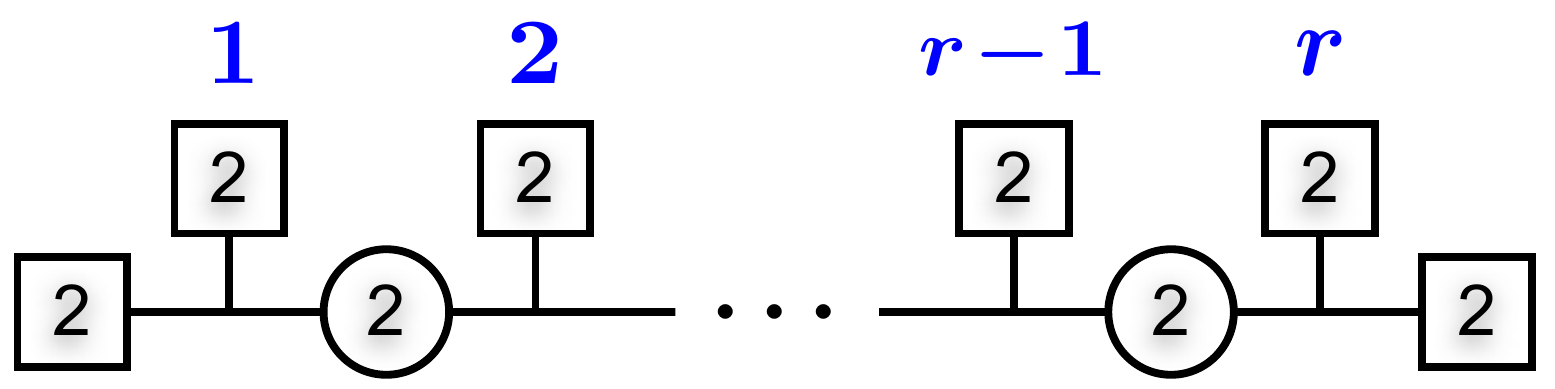}
\caption{Left: The Liouville conformal block with $(r+2)$ regular vertex operators inserted.\; Right: The quiver diagram of the $\mathcal{N}=2,SU(2)$ gauge theory corresponding the left conformal block. Each circle stands for an $SU(2)$ gauge group, and each box stands for an $SU(2)$ flavor symmetry associated with a hypermultiplet. The external momenta of the Liouville vertex operators are now encoded in the mass parameters of the hypermultiplets, while the internal momenta are related to the Coulomb branch parameters associated with the gauge groups.}
\label{fig:linear_quiver}
\end{center}
\end{figure}

In this subsection, we first review the relation between the regular conformal blocks and the Nekrasov partition functions of $SU(2)$ superconformal linear quivers \cite{AGT:2009}.
Let us consider a Liouville conformal block
\begin{eqnarray}
 \mathcal{F}_{\alpha_0}{}^{\alpha_1,\cdots,\alpha_r}{}_{\alpha_\infty}(\{\gamma_k\},\{z_k\})
\end{eqnarray}
with $(r+2)$ regular vertex operators $V_{\alpha_k} = \; :e^{2\alpha_k\phi}:$ inserted. Here $\{\gamma_k\}$ specify the intermediate channels as in the left picture of figure \ref{fig:linear_quiver} while $\{z_k\}$ denotes the loci of the vertex operator insertions.
The AGT relation states that this conformal block gives a Nekrasov partition function
\begin{eqnarray}
Z_{\rm Nek}(m_i,a_i;\epsilon_1,\epsilon_2)
\end{eqnarray}
of a $\mathcal{N}=2,SU(2)$ quiver gauge theory whose quiver diagram is given by figure \ref{fig:linear_quiver}.
Here the $\Omega$-background parameters $\epsilon_1$ and $\epsilon_2$ are related to the scale parameter $\hbar$ and the Liouville charge $Q=b+ 1/b$ through
\begin{eqnarray}
\epsilon_1 = \hbar b,\qquad \epsilon_2 = \hbar/b.
\label{eq:AGT1}
\end{eqnarray}
The mass parameters $m_i$ for the hypermultiplets are related to the external momenta in Liouville theory by
\begin{eqnarray}
\alpha_0 =\frac{m_0}{\hbar} + \frac{Q}{2},\qquad \alpha_\infty = \frac{m_\infty}{\hbar} +  \frac{Q}{2},\qquad \alpha_k = \frac{m_k}{\hbar} \quad {\rm for}\quad k=1,\cdots,r.
\label{eq:AGT2}
\end{eqnarray}
The internal momenta $\gamma_k$ are related to the Coulomb branch parameters $a_k$ by
\begin{eqnarray}
\gamma_k = \frac{a_k}{\hbar} + \frac{Q}{2} \quad {\rm for}\quad k=1,\cdots,r-1.
\end{eqnarray}
The $(r-1)$ independent loci of the Vertex operator insertions are encoded in the $(r-1)$ gauge couplings $e^{2\pi i \tau_k}$ of the quiver gauge theory. 

The Seiberg-Witten curve of the $SU(2)$ superconformal linear quiver theory is generally written as $x^2 = \phi_2(z)$ where $z$ is a coordinate on the Riemann sphere and the Seiberg-Witten differential is written as $\lambda_{\rm SW} = xdz$. This $\phi_2(z)$ is identified in \cite{AGT:2009} with the stress tensor insertion into the conformal block
\begin{eqnarray}
\phi_2(z) = \lim_{\epsilon_1,\epsilon_2\to 0}-\hbar^2\frac{\langle T(z)\,\mathcal{V}_{\alpha_0}(z_0)\cdots \mathcal{V}_{\alpha_\infty}(z_\infty)\rangle}{\langle \mathcal{V}_{\alpha_0}(z_0)\cdots \mathcal{V}_{\alpha_\infty}(z_\infty)\rangle},
\label{eq:identification_AGT}
\end{eqnarray}
where $\langle \cdots\rangle$ stands for the conformal block.

Here, we can see that the equation \eqref{eq:stress_tensor} in the matrix model is consistent with the identification \eqref{eq:identification_AGT}. In fact, \eqref{eq:stress_tensor} means that the right-hand side of \eqref{eq:identification_AGT} is given by $V'(z)^2 + f(z)$, and therefore the Seiberg-Witten curve is given by
\begin{eqnarray}
 x^2 = V'(z)^2 + f(z),
\label{eq:curve_matrix}
\end{eqnarray}
in the matrix model side.
The fact that \eqref{eq:curve_matrix} reproduces the correct Seiberg-Witten curve was already seen in \cite{Dijkgraaf:2009pc, Eguchi:2009gf}.

Since the Penner type matrix models \eqref{eq:penner} reproduces the Liouville (regular) conformal blocks, it is expected that the partition function of the Penner type matrix model reproduces the corresponding Nekrasov partition function. The parameter identifications are given by \eqref{eq:AGT1}, \eqref{eq:AGT2}, and the Coulomb branch parameters $a_k$ are identified as \cite{Dijkgraaf:2009pc, Nishinaka:2011aa}
\begin{eqnarray}
 a_k = \frac{1}{\pi i}\oint_{\mathcal{A}_k}W(z)dz - \frac{\hbar Q}{2},
\end{eqnarray}
where $\mathcal{A}_k$ is an appropriate $k$-th A-cycles of the spectral curve.

\subsection{Hitchin system with irregular singularities}

To see the effect of the colliding limits in gauge theory side, we now briefly review the six-dimensional origin of the $d=4,\mathcal{N}=2$ gauge theories. A class of $d=4, \mathcal{N}=2$ supersymmetric gauge theories, including the above mentioned $SU(2)$ superconformal quivers, is obtained by compactifying the six-dimensional $(2,0)$ supersymmetric $A_1$ theory on a Riemann surface $C$, with a partial topological twist \cite{Gaiotto:2009we, Gaiotto:2009hg}. Here the $(2,0)$ $A_1$ theory is the low energy effective theory on the stack of two M5-branes, and the topological twist leaves eight supercharges in four dimensions. We can introduce codimension two half-BPS defects on the six-dimensional theory keeping the four-dimensional $\mathcal{N}=2$ supersymmetry. Such defects are point-like on $C$ and give some punctures on it. In this paper, we only consider the case where $C$ is a (punctured) Riemann {\it sphere.} The Coulomb branch $\mathcal{B}$ of the four-dimensional theory is parameterized by the vacuum expectation values of some chiral operators, which are encoded in a quadratic differential $\phi_2$ on $C$. Here the chirality in four dimensions implies that $\phi_2$ is holomorphic on $C$.\footnote{Although this $\phi_2$ comes from the vev of some chiral operator in six dimensions, it is not a scalar but a differential on $C$, due to the topological twist to realize $\mathcal{N}=2$ supersymmetry in four dimensions.} Then, the Seiberg-Witten curve of the four-dimensional theory is written as $x^2dz^2 = \phi_2(z)$ with the Seiberg-Witten differential $xdz$.

The identification of $\mathcal{B}$ with the space of quadratic differentials is easily understood when we compactify the theory further on $S^1$ of radius $R$, which gives a $d=3,\mathcal{N}=4$ supersymmetric theory. The moduli space $\mathcal{M}$ of the three-dimensional theory is a torus fibration over $\mathcal{B}$, where the fiber directions locally parameterize the electro-magnetic Wilson loops along the $S^1$.
Here $\mathcal{M}$ can also be viewed as a moduli space of five-dimensional supersymmetric Yang-Mills theory compactified on $C$,\footnote{To be more precise, $\mathcal{M}$ is the space of five-dimensional BPS configurations which is invariant under the three-dimensional Poincare transformations.} which is the space of solutions to the Hitchin equations \cite{Gaiotto:2009hg}
\begin{eqnarray}
F + R^2 [\varphi,\overline{\varphi}] = 0,\qquad \overline{\partial}_A\varphi = \partial_A \overline{\varphi} = 0,
\end{eqnarray}
on $C$ with some boundary conditions at punctures, modulo gauge transformations. Here $F$ is a curvature of the gauge connection $A$ of a $SU(2)$-bundle $V$ on $C$, the ``Higgs field'' $\varphi$ is an $({\rm End}\,V)$-valued $(1,0)$-form on $C$, and $\partial_A = dz(\partial_{z}+A_z),\,\overline{\partial}_A = d\overline{z}(\partial_{\bar{z}} + A_{\bar{z}})$. It is known that the dimension of the Hitchin moduli space $\mathcal{M}$ is twice the dimension of $\mathcal{B}$.
The projection of $\mathcal{M}$ to $\mathcal{B}$ is given by picking a unique Casiminar ${\rm tr}(\varphi^2)$ of $\varphi$, which is identified with the quadratic differential of the four-dimensional Seiberg-Witten theory: $\phi_2 = {\rm tr}(\varphi^2)$.

The Coulomb branch $\mathcal{B}$ of the four-dimensional theory is now identified with the space of quadratic differential ${\rm tr}(\varphi^2)$ with fixed boundary conditions at punctures. Suppose that we have a puncture at $z=0$. By trivializing the bundle $V\to C$ near the puncture, the Higgs field generally has the following boundary condition near $z=0$:
\begin{eqnarray}
\varphi \sim dz\left(\frac{t_{n}}{z^{n+1}} + \frac{t_{n-1}}{z^{n}} + \cdots + \frac{t_{0}}{z} + {\rm regular}\right)
\label{eq:boundary_cond1}
\end{eqnarray}
up to gauge equivalence, where $t_k$ take values in a Cartan subalgebra of $sl(2,\mathbb{C})$. If $n=0$ then the singularity is called the ``regular singularity,'' while if $n>0$ then it is called the ``irregular singularity.'' In order to make ${\rm tr}(\varphi^2)$ singlevalued, we can consider $n\in \mathbb{N}$ or $n\in \mathbb{N}+1/2$. However, in this paper, we only consider the case of $n\in \mathbb{N}$. The boundary conditions $t_k$ specify the singular behavior of the Higgs field, and interpreted as coupling constants and masses of the four-dimensional theory. In particular the mass parameter is encoded in $t_0$. The boundary condition \eqref{eq:boundary_cond1} implies that the meromorphic quadratic differential is expanded near $z=0$ as
\begin{eqnarray}
 {\rm tr}(\varphi^2) = dz^2\left(\frac{{\rm tr}(t_n^2)}{z^{2n+2}} + \frac{2\,{\rm tr}(t_nt_{n-1})}{z^{2n+1}} + \cdots \right).
\end{eqnarray}
Note here that the coefficient of $1/z^{k+2}$ for $n \leq k\leq 2n$ are completely fixed by the boundary conditions $\{t_k\}$, while those for $k<n$ depend also on the regular terms in \eqref{eq:boundary_cond1}. This is the same situation as \eqref{eq:T_irregular} in the Liouville side and as \eqref{phi-n} in the matrix model side.

When the defects are regular, they generate hypermultiplets in four dimensions. Each such defect is characterized by a single parameter $t_0$ which is interpreted as the mass parameter of the hypermultiplet. In particular, if the $A_1$ Hitchin system on $C=\mathbb{P}^1$ has only regular singularities, then the four-dimensional theory has a weak-coupled description as a $SU(2)$ superconformal linear quiver gauge theory, as in the right picture of figure \ref{fig:linear_quiver}. The partition function of this gauge theory corresponds to a regular conformal block of the Liouville theory, via the AGT relation.

Now, let us consider the colliding limit of the Liouville conformal block which we have reviewed in section \ref{sec:Liouville}. In the Liouville side, it gives an irregular conformal block satisfying the Ward identity of the form \eqref{eq:T_irregular}. Through the identification \eqref{eq:identification_AGT}, we find that such an irregular conformal block is realized by irregular singularities in the Hitchin system on $C$. In particular, $(n+1)$ parameters $\alpha,c_1,\cdots,c_n$ of the irregular vector of rank $n$ correspond to the boundary conditions $t_0,\cdots,t_n$ of the Higgs field in the Hitchin system.\footnote{Since the Cartan subalgebra of $s\ell(2,\mathbb{C})$ is (complex) one-dimensional, each $t_k$ is determined by one complex parameter.} Such irregular singularities in the Hitchin system are known to give Argyres-Douglas theories as well as some asymptotically free $\mathcal{N}=2$ gauge theories in four dimensions \cite{Gaiotto:2009hg, G:2009, Xie:2012hs}. Then, it was pointed out \cite{Bonelli:2011aa, GT:2012} that the irregular conformal blocks of the Liouville theory should reproduce the partition function of the Argyres-Douglas type theories.

Since we have already constructed matrix models which reproduce the irregular conformal blocks of the Liouville theory, we can then provide the matrix model realization of the partition functions of the Argyres-Douglas type theories. In the next subsection we first review the Hitchin system which realizes the Argyres-Douglas theories, and in subsection \ref{subsec:matrix_for_AD} we specify the matrix models for them.

\subsection{Argyres-Douglas theories}

The Argyres-Douglas type theories were originally discovered by taking the IR limit at a special point on the Coulomb branch of $\mathcal{N}=2$ gauge theories \cite{AD:1995, Argyres:1995xn, Eguchi:1996vu, Eguchi:1996ds, Gaiotto:2010jf}. At the point on the Coulomb branch, some mutually non-local BPS particles become massless,\footnote{By ``mutually non-local'' charges, we denote electro-magnetic charges $\Gamma_1$ and $\Gamma_2$ whose Dirac-Schwinger-Zwanziger product $\langle \Gamma_1,\Gamma_2\rangle$ is non-vanishing. See also \cite{Argyres:2012fu} for a recent study on the Higgs branch of the Argyres-Douglas type theories.} which suggests the IR theory is superconformal \cite{Argyres:1995xn}. Since the theories do not have a Lagrangian description, it is not easy to perform a detailed study on them.

Recently, it was pointed out that the Hitchin system with irregular singularities realize some of the Argyres-Douglas type theories \cite{Gaiotto:2009hg, Bonelli:2011aa, GT:2012, Xie:2012hs}. Below, we briefly review the section 4 of \cite{Xie:2012hs} to describe how the $A_1$ Hitchin system realizes some class of Argyres-Douglas theories which will turn out to be related to our matrix models.

We first note that, in order for the Seiberg-Witten differential $xdz$ to have scaling dimension one, there is a constraint $[x] + [z] = 1$. Although the $SU(2)$ superconformal linear quivers realized by Hitchin system with regular singularities have $[x]=1$ and $[z]=0$, the Argyres-Douglas theories have different scaling dimensions with $[z]\neq 0$. Then the superconformal symmetry of the theories implies that there are at most two singularities on the Riemann sphere $C$, that is, one at $z=0$ and the other at $z=\infty$ \cite{Xie:2012hs}.

\subsubsection*{ $A_{2n-3}$ type Argyres-Douglas}

Let us first consider the case with a single irregular singularity of degree $n$ at $z=\infty$ and no regular singularities. In this case, the meromorphic quadratic differential $\phi_2(z)$ is written as
\begin{eqnarray}
 {\rm tr}(\varphi^2) = z^{2n-2} + u_2 z^{2n-4} + u_3 z^{2n-5} + \cdots + u_{2n-2},
\label{eq:curve_A}
\end{eqnarray}
where we rescaled $z,\varphi$ and shifted $z$ so that the coefficients of $z^{2n-2}$ and $z^{2n-3}$ are $1$ and $0$, respectively.\footnote{In this rescaling, we keep the Seiberg-Witten differential $xdz$ invariant. This is always possible if $n>0$.} Here and in the rest of this paper, we omit $dz^2$ in the right-hand side. Then the curve $x^2 = {\rm tr}(\varphi^2)$ gives the Seiberg-Witten curve of the $(A_1,A_{2n-3})$ type Argyres-Douglas theories in the notation of \cite{CNV2010, Xie:2012hs}. In this paper, we just call them $A_{2n-3}$ type Argyres-Douglas theories.

The parameters $u_k$ stand for coupling constants and the vacuum expectation values of the corresponding relevant operators of the Argyres-Douglas theories. In particular, the superconformal point is given by $u_k=0$. The variables $z$ and $x$ has dimensions $[z]=1/n,\,[x]=1-1/n$ so that $[x]+[z]=1$ and $2[x] = (2n-2)[z]$. This fixes the scaling dimensions of $u_k$ as $[u_k] = k/n$.

If $n\in \mathbb{N}$, the dimension one parameter $u_n$ is a mass deformation parameter of the theory, while $u_{2},\cdots,u_{n-1}$ of dimension less than one are coupling constants associated with some relevant operators. The remaining parameters $u_{n+1},\cdots, u_{2n-2}$ are identified with the vacuum expectation values of the relevant operators, which follows from $[u_k] + [u_{2n-k}] = 2$. On the other hand, if $n\in \mathbb{N}+1/2$ then there is no mass parameters. The parameters $u_2,\cdots,u_{n-1/2}$ are now couplings of the theory, and $u_{n+1/2},\cdots, u_{2n-2}$ are the vev of relevant operators. In this paper, we only consider $n\in \mathbb{N}$ cases.

\subsubsection*{$D_{2n}$ type Argyres-Douglas}

Now, let us consider the Hitchin system with an irregular singularity of degree $n$ at $z=0$ and a regular singularity at $z=\infty$. The meromorphic quadratic differential is now given by
\begin{eqnarray}
{\rm tr}(\varphi^2) = \frac{1}{z^{2n+2}} + \frac{u_1}{z^{2n+1}} + \cdots + \frac{u_{2n-1}}{z^3} + \frac{m^2}{z^2},
\label{eq:curve_D}
\end{eqnarray}
where we rescaled $z,\varphi$ so that the coefficient of the first term is one. We have no freedom to shift $z$ because we now have two punctures on $C$. The curve $x^2 = {\rm tr}(\varphi^2)$ is equivalent to the Seiberg-Witten curve of the $(A_1,D_{2n})$ type Argyres-Douglas theories in the notation of \cite{CNV2010, Xie:2012hs}, which we just call $D_{2n}$ type Argyres-Douglas theories. 

The scaling dimensions of $z$ and $x$ are now $[z] = -1/n, [x]= 1+1/n$, which implies $[u_k] = k/n$ and $[m]=1$. In addition to parameters $u_2,\cdots,u_{2n-2}$ which have the same interpretation as for the $A_{2n-3}$ type theories, we now have $u_1, u_{2n-1}$ and $m$. The condition $[u_1] + [u_{2n-1}] = 2$ implies that $u_1$ is a coupling constant associated with a relevant operator whose vev is given by $u_{2n-1}$. The parameter $m$ is an additional mass deformation parameter associated to the regular defect at $z=\infty$.

What is important here is that the regular puncture at $z=\infty$ is associated with a $SU(2)$ flavor symmetry. Then, we can perform ``gauging'' the diagonal $SU(2)$ flavor symmetry of $D_{2n}$ and $D_{2m}$ theories, by introducing an additional $SU(2)$ vector multiplet. Such an operation is interpreted as cutting a hole at each regular puncture and gluing them with a tube \cite{Gaiotto:2009we}, which results in a Riemann sphere with two irregular singularities at $z=0$ and $z=\infty$. Since we introduced an additional vector multiplet, the resulting gauge theory is not conformal but asymptotically free. This type of theories is called $\widehat{A}_{2m,2n}$ theory in \cite{Cecotti:2011rv, Bonelli:2011aa}.

\subsubsection*{$\widehat{A}_{2m,2n}$ theories}

Let us briefly review the Seiberg-Witten curve of the $\widehat{A}_{2m,2n}$ theory, following \cite{Bonelli:2011aa}. Suppose we have an irregular singularities of degree $m$ and $n$ at $z=0$ and $z=\infty$, respectively. Then the meromorphic quadratic differential is now written as
\begin{eqnarray}
{\rm tr}(\varphi^2) = \frac{\Lambda^2}{z^{2m+2}} + \frac{u_1}{z^{2m+1}} + \cdots + \frac{u_{2m-1}}{z^3} + \frac{u}{z^2} + \frac{\widetilde{u}_{2n-1}}{z}+ \cdots + \widetilde{u}_1 z^{2n-3}+ \Lambda^2 z^{2n-2},
\nonumber\\
\label{eq:curve_DD}
\end{eqnarray}
where we rescaled $z,\varphi$ so that the coefficients of $1/z^{2m+2}$ and $z^{2n-2}$ are the same. The Seiberg-Witten curve of the theory is given by $x^2 = {\rm tr}(\varphi^2)$. From our construction, $u_1,\cdots,u_{2m-1}$ are parameters of the $D_{2m}$ theory while $\widetilde{u}_1,\cdots,\widetilde{u}_{2n-1}$ are those of the $D_{2n}$ theory. The additional parameters $\Lambda$ and $u$ are the dynamical scale and the Coulomb branch parameter for the additional vector multiplet.

\subsubsection*{Theories involving AD theories as building blocks} 

\begin{figure}
\begin{center}
\includegraphics[width=4.5cm]{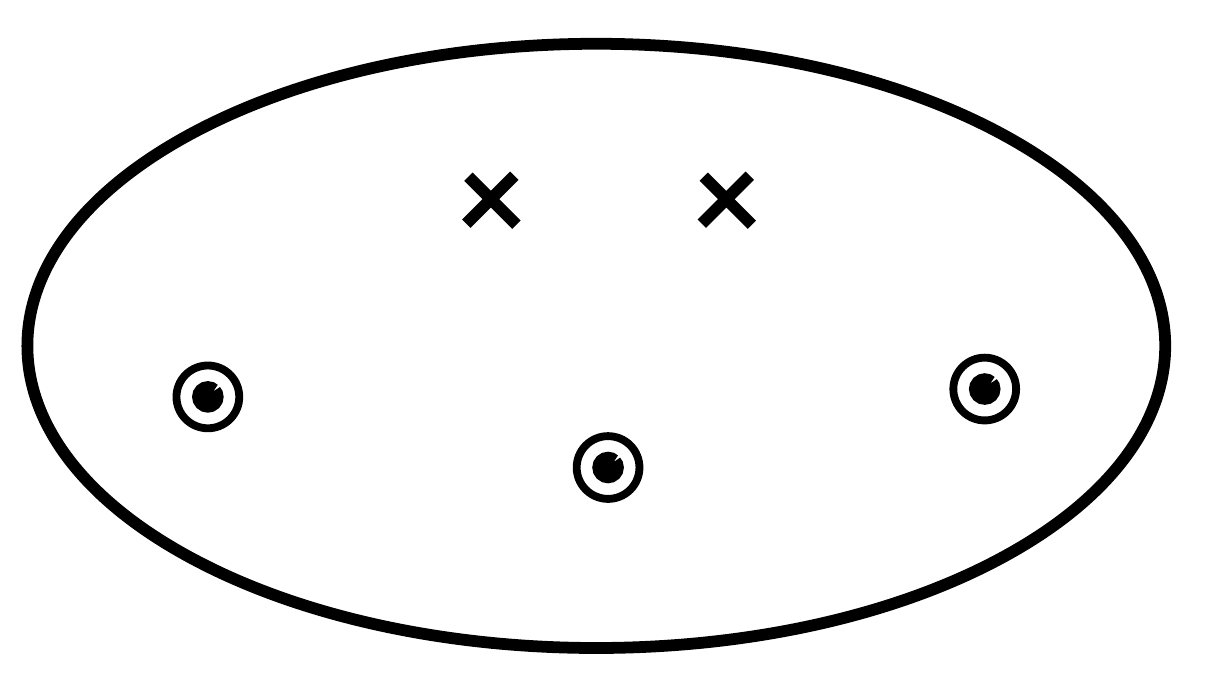}\qquad\qquad
\includegraphics[width=7cm]{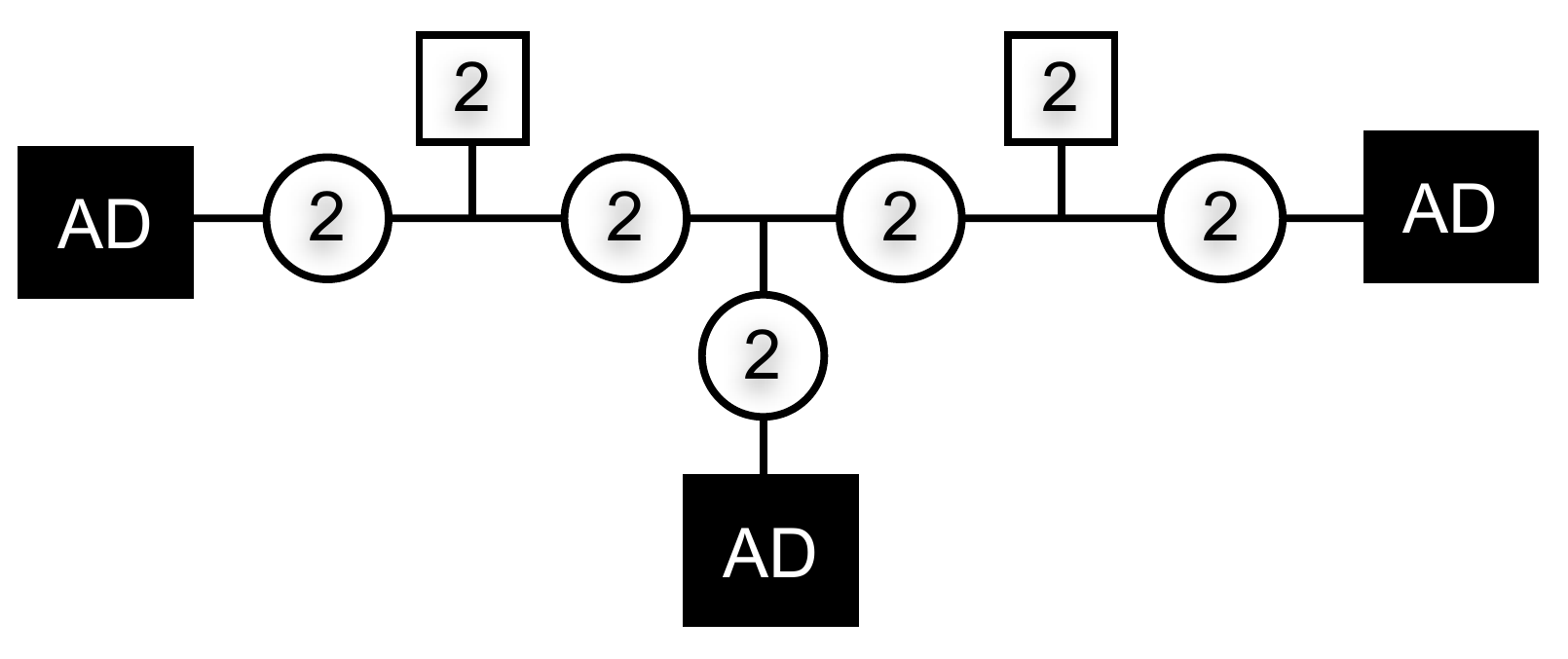}
\caption{Left: A Hitchin system on $C=\mathbb{P}^1$ with two regular and three irregular singularities. \; Right: A quiver diagram for the gauge theories associated with the Hitchin system. We here take an appropriate weak coupling limit of the marginal gauge couplings. }
\label{fig:wild_quiver}
\end{center}
\end{figure}

By generalizing the above argument, we can consider the gauge theories associated with the $A_1$ Hitchin system with many regular and irregular singularities on $C=\mathbb{P}^1$. Then, the quadratic differential has the following singular behavior near each singularity, say at $z=z_k$:
\begin{eqnarray}
{\rm tr}(\varphi^2) =  \frac{u_0^{(k)}}{(z-z_k)^{2m_k+2}} + \frac{u_1^{(k)}}{(z-z_k)^{2m_k+1}} + \cdots + \frac{u_{2m_k}^{(k)}}{(z-z_k)^2} + \cdots,
\label{eq:curve_general}
\end{eqnarray}
where $u_0^{(k)}$ denotes the dynamical scale. The parameter $u_0^{(k)},u_1^{(k)},\cdots,u_{m_k}^{(k)}$ are fixed by the boundary conditions of the Higgs field $\varphi$ at the puncture. The corresponding four-dimensional gauge theory is generically not conformal. The Seiberg-Witten curve of the theory is given by $x^2 = {\rm tr}(\varphi^2)$. The complex structure moduli space of the punctured Riemann sphere is identified with the space of marginal gauge couplings.\footnote{Note that generically not all the gauge couplings are exactly marginal.} These gauge theories are called ``wild quiver gauge theories'' in \cite{Bonelli:2011aa}, and can be constructed from $SU(2)$ vector multiplets, bifundamental and fundamental hypermultiplets, and $D_{2n}$ type Argyres-Douglas theories. For example, if there are three irregular singularities and two regular singularities on $C=\mathbb{P}^1$, then the resulting four-dimensional theory is described by a quiver diagram depicted in figure \ref{fig:wild_quiver}, in an appropriate weak-coupling regime of marginal couplings.

The dimension of the Coulomb branch of the resulting four-dimensional gauge theory is given by \cite{Bonelli:2011aa}
\begin{eqnarray}
-3 + \ell + 2i + \sum_{k = 1}^{i}\left[\frac{2m_k-1}{2}\right],
\end{eqnarray}
where $\ell$ and $i$ are the numbers of regular and irregular punctures, respectively. When all $m_k$ are integers, this reduces to
\begin{eqnarray}
-3 + (\ell + i) + \sum_{k=1}^{i}m_k.
\label{eq:dim-Coulomb}
\end{eqnarray}

\subsection{Matrix models for AD-theories}
\label{subsec:matrix_for_AD}

We now specify matrix models which realize the above mentioned gauge theories, following the conjecture that the Liouville irregular conformal blocks should reproduce the partition functions of the Argyres-Douglas type theories. Since our matrix models are associated with irregular vectors of integer ranks, we here only consider the Hitchin system with irregular singularities of integer degrees.

In particular, we explicitly write down the matrix model potentials for the $A_{2n-3}$, $D_{2n}$ Argyres-Douglas theories and $\widehat{A}_{2m,2n}$ theories for $m,n\in\mathbb{N}$. In this section, we only consider the case of $m, n\in \mathbb{N}$. We will briefly comment on the other cases $m,n\in\mathbb{N}+1/2$ in section \ref{sec:discussion}. In the matrix model side, the spectral curve coincides with the Seiberg-Witten curve of the gauge theories. The coupling constants of Argyres-Douglas theories are encoded in the matrix model potentials, while the vev of the relevant operators are encoded in the eigenvalue distributions of the matrix models. The partition function of the matrix models are conjectured to give the Nekrasov partition functions of the corresponding gauge theories.

\subsubsection*{Matrix models for $A_{2n-3}$ AD-theories}

The $A_{2n-3}$ Argyres-Douglas theory is realized by the $A_1$ Hitchin system with one irregular singularity at $z=\infty$. Such a Hitchin system is related to the one point function of an irregular vertex operator of the Liouville theory. We here only consider the case $n\in\mathbb{N}$. In the matrix model side, the corresponding potential is given by
\begin{eqnarray}
\frac{1}{\hbar}V(z) = -\sum_{k=1}^n\frac{c_kz^k}{k},
\label{eq:matrix_A}
\end{eqnarray}
 By rescaling and shifting the eigenvalues $\lambda_I$, we can set $c_n = 1$ and $c_{n-1}=0$. Such a rescaling just gives some constant multiplication to the partition function.
With the potential \eqref{eq:matrix_A}, we have
\begin{eqnarray}
f(z) \equiv 4g\sqrt{\beta}\left\langle\sum_I \frac{V'(z) - V'(\lambda_I)}{z-\lambda_I} \right\rangle = \sum_{k=0}^{n-2} d_{k}z^{k},
\end{eqnarray}
with
\begin{eqnarray}
d_{k} = -2\hbar^2 b\sum_{I=1}^{N}\left\langle\sum_{\ell=k}^{n-2}c_{\ell + 2} \lambda_I^{\ell - k} \right\rangle.
\end{eqnarray}
Then, the spectral curve of the matrix model $x^2 = V'(z)^2 + f(z)$ is exactly the same form as \eqref{eq:curve_A}. In particular, the coefficients $u_2,\cdots,u_{n-1}$ in \eqref{eq:curve_A} is completely determined by the $(n-2)$ coupling constants in the potential \eqref{eq:matrix_A}.\footnote{Recall that we are now setting $c_n=1$ and $c_{n-1}=0$.} The mass deformation parameter $u_{n}$ of the AD theory is determined by
\begin{eqnarray}
d_{n-2} = -2\hbar^2 bN.
\end{eqnarray}
On the other hand, the other coefficients $u_{n},u_{n+1},\cdots,u_{2n-2}$ depend on $d_2,\cdots,d_{n-1}$. In fact, these $(n-2)$ quantities are determined by $(n-2)$ {\it filling fractions} of the matrix model. The matrix model originally have $N$ eigenvalues, and they are distributed along the cuts of the spectral curve. Now, the matrix model has $(n-1)$ cuts, and therefore we have $(n-2)$ independent filling fractions:
\begin{eqnarray}
g\sqrt{\beta}N_k = \frac{1}{2\pi i}\oint_{\mathcal{A}_k}W(z)dz,
\end{eqnarray}
where $\mathcal{A}_k$ is a cycle encircling the $k$-th cut and we have a constraint $\sum_{k=1}^{n-1}N_k = N$. Then, fixing all the filling fractions determines $\{d_k\}$.

Thus, we have found that the couplings and mass parameter of the AD theory is encoded in the couplings and the matrix size $N$ of the matrix model. Since the matrix size $N$ is related to the Liouville momenta $\alpha_\infty$ through $\alpha_\infty + bN = Q$, this is consistent with the fact that the mass deformation parameter of the AD theory comes from $t_0$ in \eqref{eq:boundary_cond1} in the Hitchin system. We have also found that the vacuum expectation values of the relevant operators in the AD theory are now encoded in the filling fractions of the matrix model.

We here briefly note that \eqref{eq:matrix_A} is consistent with the original argument in \cite{Dijkgraaf:2002fc}. In fact, the $A_{2n-3}$ type Argyres-Douglas theories can be geometrically engineered by type IIB string theory on a Calabi-Yau singularity
\begin{eqnarray}
uv + x^2 + z^{2n-2} = 0,
\end{eqnarray}
which has the matrix model realization with the potential $V(z) = z^{n}/n$. Including the relevant deformations of the gauge theory now corresponds to deforming the potential as \eqref{eq:matrix_A}. In this paper, we instead derive \eqref{eq:matrix_A} through the scaling limit of the Penner type matrix models which has a direct connection to the Nekrasov partition function of $SU(2)$ superconformal linear quivers.

\subsubsection*{Matrix models for $D_{2n}$ AD-theories}

We now turn to the $D_{2n}$ type Argyres-Douglas theories. The corresponding Hitchin system has an irregular singularity of degree $n$ at $z=0$ and a regular singularity at $z=\infty$. In the Liouville side, this setup corresponds to considering an inner product $\langle R_1|I^{(n)}\rangle$. We again concentrate on the case of $n\in\mathbb{N}$. Then, the corresponding matrix model has the potential \eqref{eq:pot1}, where we set $c_n=1$ by rescaling the eigenvalues $\lambda_I$.
Then the quantity $f(z)$ is now written as
\begin{eqnarray}
f(z) = \sum_{k=0}^{n-1}\frac{d_k}{z^{k+2}},
\end{eqnarray}
with $d_k = -\hbar^2 v_k(\log Z_{M})$ in the notation of \eqref{eq:V-f}. The term proportional to $1/z$ vanishes due to the symmetry $\langle \sum_IV'(\lambda_I)\rangle = 0$.

The spectral curve of the matrix model $x^2 = V'(z)^2 + f(z)$ is now of the same form as \eqref{eq:curve_D}. The couplings $u_1,\cdots,u_{n-1}$ and mass parameter $u_n$ of the AD theory are completely fixed by $n$ coupling constants in the matrix model potential \eqref{eq:pot1}.\footnote{Recall that we are now setting $c_n=1$.} The mass parameter $m^2$ associated with the regular singularity is now determined by $d_0$. In fact, this $d_0$ is fixed by the matrix size $N$. To see this, we recall the loop equation
\begin{eqnarray}
g^2W(z,z) + W(z)^2 + \frac{\hbar Q}{2}W'(z) + V'(z)W(z) - \frac{f(z)}{4} = 0.
\end{eqnarray}
Since $W(z) = \hbar bN/(2z) + \mathcal{O}(z^{-2}),\, W(z,z) = \mathcal{O}(z^{-4}),\, V'(z) = \hbar\alpha/z + \mathcal{O}(z^{-2})$ around infinity, we obtain
\begin{eqnarray}
d_0 = \hbar^2 (b^2N^2 -  QbN + 2\alpha bN)
\end{eqnarray}
without any approximations. Thus, the mass parameter $m^2$ in \eqref{eq:curve_D} is encoded in $N$ in the matrix model. Since $N$ is related to $\alpha_\infty$ through $\alpha + \alpha_\infty + bN = Q$, this is consistent with the fact that $m^2$ in \eqref{eq:curve_D} comes from the simple pole of the Higgs field in \eqref{eq:boundary_cond1}.

On the other hand, the vacuum expectation values of the relevant operators $u_{n+1},\cdots,u_{2n-1}$ are determined by the remaining $d_k$ for $k=1,\cdots,n-1$. The $(n-1)$ quantities are determined by $(n-1)$ filling fractions of the matrix model. In fact, since the spectral curve is of the form
\begin{eqnarray}
x^2 = \frac{\mathcal{P}_{2n}(z)}{z^{2n+2}}
\end{eqnarray}
with some $2n$-th order polynomial $\mathcal{P}_{2n}(z)$, 
the matrix model has now $n$ cuts. Therefore, there are $(n-1)$ independent filling fractions
\begin{eqnarray}
g\sqrt{\beta}N_k = \frac{1}{2\pi i}\oint_{\mathcal{A}_k}W(z)dz.
\end{eqnarray}

Thus, we have again found that the couplings and mass parameter of the AD theory is encoded in the couplings and the matrix size $N$ of the matrix model, while the vev of the relevant operators in the AD theory are encoded in the filling fractions of the matrix model.

\subsubsection*{Matrix models for $\widehat{A}_{2m,2n}$ theories}

The $\widehat{A}_{2m,2n}$ theory is associated to the Hitchin system with two irregular singularities of degree $m$ and $n$ at $z=0$ and $z=\infty$, respectively. We here assume $m,n\in\mathbb{N}$ and $m,n\geq 2$. In the Liouville side, the partition function of this gauge theory is expected to give an inner product $\langle I^{(m)}|I^{(n)}\rangle$.\footnote{Note here that the $\widehat{A}_{2,2}$ theory is the $SU(2)$ gauge theory with two flavors, and the matrix model realization for this theory was studied in \cite{Eguchi:2009gf}.}
 The corresponding matrix model has the potential \eqref{eq:general2}, where we set $c_m^{(0)}=c_n^{(\infty)} = \Lambda$ by rescaling eigenvalues $\lambda_I$. Then the quantity $f(z)$ is now written as
\begin{eqnarray}
f(z) = \sum_{k=-n}^{m-1}\frac{d_k}{z^{k+2}},
\end{eqnarray}
where
\begin{eqnarray}
d_k &=& -2\hbar^2 b \sum_I^N\left\langle \sum_{\ell = k+1}^{m}\frac{c_{\ell}^{(0)}}{\lambda_I^{\ell-k}}\right\rangle\qquad {\rm for}\qquad -1 \leq k\leq m-1,
\\
d_k &=& -2\hbar^2 b\sum_{I=1}^N\left\langle \sum_{\ell=|k|}^{n}c_{\ell}^{(\infty)}\lambda_I^{\ell+k}\right\rangle \qquad {\rm for}\qquad -n \leq k \leq -2.
\end{eqnarray}

The spectral curve of the matrix model $x^2=V'(z)^2 + f(z)$ is then exactly of the same form as \eqref{eq:curve_DD}. The couplings and mass parameters $u_1,\cdots,u_m$ of the $D_{2m}$ theory are completely fixed by the couplings $\alpha^{(0)},c_1^{(0)},\cdots,c_{m-1}^{(0)}$ of the matrix model, while the couplings $\widetilde{u}_1,\cdots,\widetilde{u}_{n-1}$ of the $D_{2n}$ theory are fixed by $c_1^{(\infty)},\cdots,c_{n-1}^{(\infty)}$ in the potential \eqref{eq:general2}. The mass deformation parameter $\widetilde{u}_n$ of the $D_{2n}$ theory is encoded in $d_{-n} = -2\hbar^2\Lambda bN$, that is, the matrix size $N$. The vev of the relevant operators in the $D_{2m}$ and $D_{2n}$ theories are respectively encoded in $d_1,\cdots,d_{m-1}$ and $d_{-1},\cdots,d_{-n+1}$, while the Coulomb moduli $u$ for the additional vector multiplet is determined by $d_0$. These $(n+m-1)$ moduli parameters are fixed by $(n+m-1)$ filling fractions of the matrix model. In fact, since the spectral curve is written in the form
\begin{eqnarray}
x^2 = \frac{\mathcal{P}_{2m+2n}(z)}{z^{2m-2}}
\end{eqnarray}
with some $(2m+2n)$-th order polynomial $\mathcal{P}_{2m+2n}(z)$, the matrix model has now $(m+n-1)$ independent filling fractions:
\begin{eqnarray}
g\sqrt{\beta}N_k = \frac{1}{2\pi i}\oint_{\mathcal{A}_k} W(z)dz.
\end{eqnarray}

\subsubsection*{Matrix models for wild quivers}

When we consider an $A_1$-Hitchin system with many regular and irregular singularities, the four-dimensional gauge theory is generically an asymptotically free theory involving Argyres-Douglas theories as building blocks. The Seiberg-Witten curve of the gauge theory is written in the form \eqref{eq:curve_general}. The corresponding matrix model which gives a general irregular conformal block is described by the potential
\begin{eqnarray}
\frac{1}{\hbar}V(z) = \sum_{k=1}^{r}\left(\alpha^{(k)}\log(z-z_k) - \sum_{j=1}^{m_k}\frac{c_j^{(k)}}{j(z-z_k)^j}\right) - \sum_{k=1}^{m_\infty}\frac{c_k^{(\infty)}z^k}{k},
\end{eqnarray}
where we assume $c_{m_k}^{(k)}\neq 0$ if $m_k> 0$. The singularity at $z=z_k$ is regular if $m_k=0$ while it is irregular if $m_k>0$. Note here that we always have a (regular or irregular) singularity at $z=\infty$ unless the size of the matrix $N$ vanishes.\footnote{This is due to the fact that the resolvent is generally expanded around $z=\infty$ as $W(z) = \hbar b N/(2z) + \mathcal{O}(z^{-2})$.}

The spectral curve $x^2 = V'(z)^2 + f(z)$ of this matrix model is generally of the form
\begin{eqnarray}
x^2 = \frac{\mathcal{P}_{d}(z)}{\prod_{k=1}^{r}(z-z_k)^{2m_k+2}},
\label{eq:curve1}
\end{eqnarray}
where $d=2r + \sum_{k=1}^{r}2m_k + 2m_\infty - 2$ and $\mathcal{P}_d(z)$ is a $d$-th order polynomial of $z$. This is easily shown when $m_\infty >0$. If $m_\infty=0$, we can see this as follows. First we note that
\begin{eqnarray}
\frac{1}{\hbar^2}V'(z)^2 &=& \left(\sum_{k=1}^r \frac{c_{m_k}^{(k)}}{(z-z_k)^{m_k+1}} + \frac{c_{m_k-1}^{(k)}}{(z-z_k)^{m_k}} + \cdots + \frac{\alpha^{(k)}}{z-z_k}\right)^2,
\nonumber
\\
f(z) &=& \sum_{k=1}^r \left(\frac{d_{m_k-1}^{(k)}}{(z-z_k)^{m_k+1}} + \frac{d_{m_k-2}^{(k)}}{(z-z_k)^{m_k}} + \cdots + \frac{d_{-1}^{(k)}}{(z-z_k)}\right),
\end{eqnarray}
for some coefficients $d_j^{(k)}$. Then, one might think that the spectral curve $x^2 = V'(z)^2 + f(z)$ is of the form
\begin{eqnarray}
x^2 = \frac{\mathcal{P}_{d+1}(z)}{\prod_{k=1}^{r}(z-z_k)^{2m_k+2}},
\label{eq:curve2}
\end{eqnarray}
where $\mathcal{P}_{d+1}(z)$ is a $(2r + \sum_{k=1}^r 2m_k -1)$-th order polynomial of $z$. However, the coefficient of $z^{d+1}$ in the numerator of \eqref{eq:curve2} is 
\begin{eqnarray}
\sum_{k=1}^r d_{-1}^{(k)},
\end{eqnarray}
which turns out to vanish. In fact, this quantity is the residue of $f(z)$ at infinity and equivalent to
\begin{eqnarray}
-2\hbar b\left\langle \sum_{I=1}^N V'(\lambda_I)\right\rangle = 0
\end{eqnarray}
when $m_\infty = 0$. Thus, the spectral curve is of the form \eqref{eq:curve1} rather than \eqref{eq:curve2} even when $m_\infty = 0$. This fact implies that the number of independent filling fractions of the matrix model is always
\begin{eqnarray}
\frac{d}{2}-1 &=& -3 + (r+1) + \sum_{k=1}^rm_k + m_\infty.
\end{eqnarray}
Since $(r+1)$ is the total number of (regular and irregular) singularities,\footnote{Note that we always have a (regular or irregular) singularity at infinity.} this is exactly the same as the dimension of the Coulomb branch of the corresponding gauge theory \eqref{eq:dim-Coulomb}.

\subsubsection*{Partition functions of matrix models and gauge theories}

We have seen that our matrix models for irregular conformal blocks correctly reproduce the Seiberg-Witten curves of some Argyres-Douglas theories and wild quiver gauge theories. Recalling that the partition functions of the Penner type matrix models are conjectured to reproduce the Nekrasov partition functions of the $SU(2)$ linear quivers, we now conjecture that the partition functions of our matrix models reproduce the Nekrasov partition functions of the corresponding $A_{2n-3}, D_{2n}$-type Argyres-Douglas theories and wild quiver gauge theories. The parameters are identified as \eqref{eq:AGT1} and 
\begin{eqnarray}
&&\alpha^{(0)} = \frac{m_0}{\hbar} + \frac{Q}{2},\qquad \alpha^{(\infty)} = \frac{m_\infty}{\hbar} + \frac{Q}{2},\qquad \alpha^{(k)} = \frac{m_k}{\hbar},
\end{eqnarray}
where $m_0,m_\infty,m_k$ stand for some mass parameters of the gauge theory. The Coulomb branch parameters are identified as
\begin{eqnarray}
a_k = \frac{1}{\pi i}\oint_{\mathcal{A}_k}W(z)dz + \frac{\hbar Q}{2},\qquad (a_D)_k = \frac{1}{\pi i}\oint_{\mathcal{B}_k}W(z)dz + \frac{\hbar Q}{2},
\end{eqnarray}
where we take the 1-cycles of the spectral curve so that their intersections are given by
\begin{eqnarray}
\left\langle\mathcal{A}_j,\mathcal{B}_k\right\rangle = \delta_{jk},\qquad \left\langle \mathcal{A}_j,\mathcal{A}_k\right\rangle = \left\langle \mathcal{B}_j,\mathcal{B}_k\right\rangle = 0.
\end{eqnarray}

In the genus expansion of our matrix model
\begin{eqnarray}
\log Z_M = \sum_{n=0}^\infty F_n (i\hbar)^{n-2},
\label{eq:genus-exp}
\end{eqnarray}
the leading term $F_0$ should particularly give the prepotential of the corresponding gauge theory. In fact, from the general property of the matrix model, it follows that $F_0$ satisfies the special geometry relation:
\begin{eqnarray}
\frac{\partial F_0}{\partial a_k} = (a_D)_k,
\end{eqnarray}
which is necessary in the gauge theory side. Thanks to this relations, $F_0$ is determined by the spectral curve $x^2 = \phi_2(z)$ and the meromorphic one-form $xdz = (2W_0+V')dz$. Since we have already checked that the spectral curve of our matrix models coincides with the Seiberg-Witten curve of the corresponding gauge theories, at least we can see that $F_0$ correctly describes the IR physics of the corresponding gauge theories. It is worth studying the higher order terms of \eqref{eq:genus-exp} further.

\section{Summary and discussions}
\label{sec:discussion}

In this paper, we have constructed matrix models which reproduce the irregular conformal blocks of the Liouville theory on sphere. We have studied the matrix side of the colliding limit of the Liouville vertex operators, and pointed out that if the matrix model potential is written as a sum of logarithmic and/or rational functions then its partition function reproduces a conformal block with insertions of regular and/or irregular states of the Liouville theory on sphere. In section \ref{sec:irregular_partition}, we have particularly studied the $D_{2n}$-type matrix model in detail, and show that the partition function of the matrix model correctly reproduces the inner product of a regular and an irregular states. We have also shown that our matrix models generally reproduce the small $c_2$ behavior of the irregular state $|I^{(2)}(c_1,c_2;\alpha)\rangle$ proposed in \cite{GT:2012}. In section \ref{sec:gauge_theories}, we have also discussed the relation between our matrix models and the Argyres-Douglas theories in four dimensions. We have shown that our matrix models correctly reproduce the Seiberg-Witten curves of the corresponding gauge theories.

We should here mention that we have not studied irregular singularities of half-integer degree in this paper. Especially, we have not studied matrix models for $A_{2n}$ or $D_{2n+1}$-type Argyres-Douglas theory. In fact, these theories cannot be realized by logarithmic or rational potentials of the matrix model. For example, the Seiberg-Witten curve of the $A_{2n}$ Argyres-Douglas theory is of the form
\begin{eqnarray}
x^2 = z^{2n+1} + u_2z^{2n-1} + \cdots + u_{2n+1},
\end{eqnarray}
and if this is realized as the (planar) spectral curve of the matrix model $x^2 = V'(z)^2 + f(z)$ then it seems likely that the potential $V(z)$ should have a square-root term $z^{n+1/2}$. However, it is not straightforward to generalize our argument to such a square-root potential. This complication comes from a different singular behavior near the irregular singularity of half-integer degree, which needs to be studied further. Note also that the colliding limits of Liouville vertex operators have not yet been well-established for irregular singularities of half-integer degrees.

For future works, it would be interesting to study the higher orders of the $c_2$-expansion \eqref{eq:ansatz} in the matrix model side, which will lead to a matrix model expression of an inner product $\langle G| I^{(n)}_{2k}\rangle$ with a generalized descendant $|I^{(n)}_{2k}\rangle$, and to extend the method to the half-integer rank case.

The application of our matrix models to the quantization problem of Hitchin system is also an interesting future problem. In \cite{Bonelli:2011na}, the Penner type matrix models were used to quantize the Hitchin system with regular singularities. Since our matrix models can take into account irregular singularities in the Hitchin system, it would be interesting to generalize the argument in \cite{Bonelli:2011na} by using our matrix models.

It is also worth studying the generalization to the higher rank of gauge groups. As pointed out in \cite{Dijkgraaf:2009pc}, the higher rank gauge groups correspond to the $\beta$-ensemble of multi matrix models. By generalizing our argument to the multi matrix models, we can construct matrix models for $A_{n}$ Hitchin system with irregular singularities. Such a generalization will give a matrix model realization of $(A_n,A_m)$-type Argyres-Douglas theories in the notation of \cite{CNV2010}.

\subsection*{Acknowledgments} 
We would like to thank Goro Ishiki for illuminating discussions and important comments.
This work is partially supported by the National Research Foundation of Korea (NRF) grant funded by the Korea government (MEST) 2005-0049409.

\end{document}